\documentclass[10pt, conference, onecolumn,  compsocconf ]{IEEEtran}

\usepackage[cmex10]{amsmath}
\usepackage{amsthm, amssymb}
\usepackage{cite}
\usepackage{array}
\usepackage{graphicx}
\usepackage[dvipsnames,usenames]{color}
\usepackage{xspace}

\def\doctype{1}

\def\tsubmission{2}
\ifnum\doctype=\tsubmission
	\newcommand{\full}[1]{}
	\newcommand{\submit}[1]{#1}
\else
	\newcommand{\full}[1]{#1}
	\newcommand{\submit}[1]{}
\fi

\newtheorem{theorem}{Theorem}
\newtheorem{lemma}[theorem]{Lemma}

\theoremstyle{definition}
\newtheorem{definition}{Definition}
\newtheorem{corollary}[theorem]{Corollary}

\newtheorem{fact}[theorem]{Fact}

\newcommand{\vs}{{\sc vs}\xspace}
\newcommand{\vslpi}{{\sc vsall-i}\xspace}
\newcommand{\vslpu}{{\sc vsany-u}\xspace}
\newcommand{\vsmax}{{\sc vsmax}\xspace}
\newcommand{\vsmaxi}{{\sc vsmax-i}\xspace}
\newcommand{\vsmaxu}{{\sc vsmax-u}\xspace}

\newcommand{\vbp}{{\sc vbp}\xspace}
\newcommand{\vc}{{\sc vc}\xspace}

        {\hspace*{\fill}$\Box$\par}

\begin{document}

\title{Tight Bounds for Online Vector Scheduling}
\author{\IEEEauthorblockN{Sungjin Im\IEEEauthorrefmark{1}\hspace*{20pt} Nathaniel Kell\IEEEauthorrefmark{2}\hspace*{20pt} Janardhan Kulkarni\IEEEauthorrefmark{3}\hspace*{20pt} Debmalya Panigrahi\IEEEauthorrefmark{2}}
\vspace*{20pt}
\IEEEauthorblockA{\IEEEauthorrefmark{1}Electrical Engineering and Computer Science,
University of California at Merced,
Merced, CA, USA.\\
{\tt Email: sim3@ucmerced.edu}}
\IEEEauthorblockA{\IEEEauthorrefmark{2}Department of Computer Science,
Duke University,
Durham, NC, USA.\\
{\tt Email: \{kell,debmalya\}@cs.duke.edu}}
\IEEEauthorblockA{\IEEEauthorrefmark{3}Microsoft Research,
Redmond, WA, USA.\\
{\tt Email: jakul@microsoft.com}}
}

\maketitle

\begin{abstract}
Modern data centers face a key challenge of effectively serving user requests that arrive online. 
Such requests are inherently \emph{multi-dimensional} and characterized by demand vectors 
over multiple resources such as processor cycles, storage space, and network bandwidth. Typically,
different resources require different objectives to be optimized, and $L_r$ norms of loads are among 
the most popular objectives considered. Furthermore, the server clusters are also often heterogeneous 
making the scheduling problem more challenging.

To address these problems, we consider the {\em online vector scheduling} problem in this paper.
Introduced by Chekuri and Khanna (SIAM J. of Comp. 2006), vector scheduling is a generalization of
classical load balancing, where every job has a vector load instead of a scalar load. The scalar 
problem, introduced by Graham in 1966, and its many variants (identical and unrelated machines, 
makespan and $L_r$-norm optimization, offline and online jobs, etc.) have been extensively studied 
over the last 50 years. In this paper, we resolve the online complexity of the 
vector scheduling problem and its important generalizations --- for all $L_r$ norms and in both the 
identical and unrelated machines settings. Our main results are:

\begin{itemize}
	\item For {\bf identical machines},
	we show that the optimal competitive ratio is $\Theta(\log d / \log \log d)$
	by giving an online lower bound and an algorithm with an asymptotically 
	matching competitive ratio. The lower bound is technically challenging,
	and is obtained via an online lower bound for the minimum mono-chromatic
	clique problem using a novel online coloring game and randomized coding 
	scheme. Our techniques also extend to asymptotically 
	tight upper and lower bounds for general $L_r$ norms. 
	\item For {\bf unrelated machines},
	we show that the optimal competitive ratio is $\Theta(\log m + \log d)$
	by giving an online lower bound that matches a previously known upper
	bound. Unlike identical machines, however, extending these results, 
	particularly the upper bound, to general $L_r$ norms requires new ideas.
	In particular, we use a carefully constructed potential function that 
	balances the individual $L_r$ objectives with the overall (convexified) 
	min-max objective to guide the online algorithm and track the changes
	in potential to bound the competitive ratio.
\end{itemize}
\end{abstract}


\section{Introduction}
\label{sec:intro} 

A key algorithmic challenge in modern data centers is the scheduling of 
online resource requests on the available hardware. Such requests are 
inherently {\em multi-dimensional} and simultaneously ask for multiple resources
such as processor cycles, network bandwidth, and storage 
space~\cite{ghodsi2011dominant, lee2011heterogeneity, GrandlAKRA14} 
(see also multi-dimensional load balancing in 
virtualization~\cite{vmware, jung2008generating}). In addition
to the multi-dimensionality of resource requests, another challenge is the
{\em heterogeneity} of server clusters because 
of incremental hardware deployment and the use of dedicated specialized hardware 
for particular tasks~\cite{ahmad2012tarazu, ghodsi2013choosy, zhang2013energy}. 
As a third source of non-uniformity, the objective of the load balancing 
exercise is often defined by the application at hand and the resource being 
allocated. In addition to the traditional goals of minimizing maximum 
($L_{\infty}$ norm) and total ($L_1$ norm) machine loads, 
various intermediate $L_r$ norms
are also important for specific applications. For example, the $L_2$ norm
of machine loads is suitable for disk storage~\cite{ChandraW75,CodyC76}
while the $L_r$ norm for $r$ between 2 and 3 is used for modeling energy
consumption~\cite{Pelley09, Albers10, YaoDS95}.

In the algorithmic literature, the (single dimensional) load balancing 
problem, also called list scheduling, has a long history since the 
pioneering work of Graham in 1966~\cite{Graham66}. However, the 
multi-dimensional problem, introduced by Chekuri and Khanna~\cite{ChekuriK04}
and called {\em vector scheduling} (\vs), remains less understood.
In the simplest version of this problem, each job has a vector load 
and the goal is to assign the jobs to machines so as to minimize the 
maximum machine load over all dimensions. 
As an example of our limited understanding of this problem, 
we note that the approximation complexity of this most basic version is 
not resolved yet --- the current best approximation factor is 
$O(\log d/\log \log d)$ (e.g.,~\cite{HarrisS13}), where $d$ is the number 
of dimensions, while only an $\omega(1)$ lower bound is known~\cite{ChekuriK04}.
In this paper, we consider the online version of this problem, i.e., 
where the jobs appear in a sequence and have to be assigned irrevocably to a machine
on arrival. Note that this is the most common scenario in the data center
applications that we described earlier, and in other real world settings. 
In addition to the basic setting described above, we also 
consider more general scenarios to capture the practical challenges that
we outlined. In particular, we consider this problem in both the 
identical and unrelated machines settings, the latter capturing the 
non-uniformity of servers. Furthermore, we also consider all $L_r$ norm 
objectives of machine loads in addition to the makespan ($L_{\infty}$) 
objective \footnote{Our $L_r$-norms are typically referred to as $p$-norms or $L_p$- norms. We use $L_r$-norms to reserve the letter $p$ for job processing times.}. In this paper, we completely resolve the online complexity of 
all these variants of the vector scheduling problem.

Formally, 
there are $n$ jobs (denoted $J$) that arrive online and must
be immediately and irrevocably assigned on arrival
to one among a fixed set of $m$ machines (denoted $M$). We denote the
$d$-dimensional load vector of job $j$ on machine $i$ by 
$p_{i,j} = \langle p_{i,j}(k): k\in [d]\rangle$, which is revealed 
on its online arrival. For identical machines, the load of 
job $j$ in dimension $k$ is identical for all machines $i$, and
we denote it $p_j(k)$. Let us denote the assignment 
function of jobs to machines by $f: J \rightarrow M$. 
An assignment $f$ produces a {\em load} of 
$\Lambda_{i}(k) = \sum_{j:f(j) =i}p_{i,j}(k)$ in dimension $k$
of machine $i$; we succinctly denote the machine loads in 
dimension $k$ by an $m$-dimensional vector $\Lambda(k)$. 
(Note that for the scalar problem, there is only one such machine 
load vector.) 

\medskip \noindent \textbf{The makespan norm.}
We assume (by scaling) that the optimal makespan norm on each 
dimension is 1. Then, the \vs problem for the makespan norm (denoted \vsmax) 
is defined as follows.
\begin{definition} \label{def:makespan} {\vsmax}: For any dimension $k$, the objective is
the maximum load over all machines, i.e.,
\begin{equation*}
	\|\Lambda(k)\|_{\infty} =  \max_{i \in M}\Lambda_{i}(k). 
\end{equation*} 
\end{definition}
\noindent
An algorithm is said to be $\alpha$-competitive if 
$\|\Lambda(k)\|_{\infty} \leq \alpha$ for every dimension $k$.
We consider this problem in both the {\em identical machines} (denoted \vsmaxi) 
and the {\em unrelated machines} (denoted \vsmaxu) settings. First, we state 
our result for identical machines. 
\begin{theorem} 
\label{thm:vsmaxi} 
	There is a lower bound of $\Omega\left(\frac{\log d}{\log \log d}\right)$ on the 
	competitive ratio of online algorithms for the \vsmaxi problem. Moreover, 
	there is an online algorithm whose competitive ratio asymptotically 
	matches this lower bound.
\end{theorem} 
The upper bound is a slight improvement over the previous best 
$O(\log d)$~\cite{AzarCKS13,MeyersonRT13}, but the only lower bound known
previously was NP-hardness of obtaining an $O(1)$-approximation for the
offline problem~\cite{ChekuriK04}. We remark that while the offline 
approximability remains unresolved, the best offline algorithms currently 
known (\hspace{-1sp}\cite{AzarCKS13,MeyersonRT13}, this paper) are in fact online. 
Also, our lower bound is information-theoretic, i.e., relies on the 
online model instead of computational limitations. 

For unrelated machines (\vsmaxu), an $O(\log m + \log d)$-competitive
algorithm was given by Meyerson~{\em et al.}~\cite{MeyersonRT13}.
We show that this is the best possible. 
\begin{theorem}
\label{thm:vsmaxu}
	There is a lower bound of $\Omega(\log m + \log d)$ on the 
	competitive ratio of online algorithms for the \vsmaxu problem.
\end{theorem}

\noindent \textbf{Extensions to other $L_r$ norms.}
As we briefly discussed above, there are many applications where an 
$L_r$ norm (for some $r\geq 1$) is more suitable than the makespan 
norm. First, we consider identical machines, and aim to simultaneously 
optimize {\em all norms} on all dimensions (denoted \vslpi).
\begin{definition}
{\vslpi}: For dimension $k$ and norm $L_r$, $r \geq 1$, the objective is
\begin{equation*}
	\|\Lambda(k)\|_r =  \Big(\sum_{i \in M}\Lambda_{i}^r(k)\Big)^{1/r}. 
\end{equation*} 
\end{definition}
\noindent
An algorithm is said to $\alpha_r$-competitive for the 
$L_r$ norm if $\|\Lambda(k)\|_r \leq \alpha_r$ 
for every dimension $k$ and every $L_r$ norm, $r \geq 1$.
The next theorem extends Theorem~\ref{thm:vsmaxi} to an all norms 
optimization.
\begin{theorem}
\label{thm:vsall}
	There is an online algorithm for the \vslpi problem that
	obtains a competitive ratio of 
	$O\Big(\Big(\frac{\log d}{\log \log d}\Big)^\frac{r-1}{r}\Big)$, 
	simultaneously for all $L_r$ norms. Moreover, these competitive ratios
	are tight, i.e., there is a matching lower bound for every individual 
	$L_r$ norm.
\end{theorem}

For unrelated machines, there is a polynomial lower bound 
for simultaneously optimizing multiple $L_r$ norms, even
with scalar loads.
This rules out an {\em all norms} approximation. Therefore,
we focus on an {\em any norm} approximation, where 
the algorithm is given norms $r_1, r_2, \ldots, r_d$
(where $1 \leq r_k \leq \log m$),\footnote{For any  $m$-dimensional vector $x$,
$\| x \|_\infty = \Theta(\| x \|_{\log m})$. 
Therefore, for any $r_k > \log m$, an algorithm can instead use a $L_{\log m}$
norm to approximate an $L_{r_k}$ norm objective up to constant distortion. Thus, 
in both our upper and lower bound results we restrict $1 \leq r_k \leq \log m$.}
and the goal is to minimize the $L_{r_k}$ norm 
for dimension $k$. The same lower bound also rules out
the possibility of the algorithm being competitive against
the optimal value of each individual norm in their 
respective dimensions. 
We use a standard 
trick in multi-objective optimization to circumvent 
this impossibility: we only require the algorithm to be 
competitive against any given feasible {\em target} vector
$T = \langle T_1, \ldots, T_d \rangle$. For ease of notation, 
we assume wlog (by scaling) that $T_k = 1$ for all dimensions
$k$.\footnote{A target vector is feasible if there is an assignment 
such that for every
dimension $k$, the value of  the $L_{r_k}$ norm in that dimension 
is at most $T_k$. Our results do not
rely heavily on the exact feasibility of the target vector; if 
there is a feasible solution that violates targets in all dimensions
by at most a factor of $\beta$, then our results hold with 
an additional factor of $\beta$ in the competitive ratio.}
Now, we are ready to define the \vs problem with arbitrary 
$L_r$ norms for unrelated machines --- we call this problem \vslpu.
\begin{definition}
{\vslpu}: For dimension $k$, the objective is
\begin{equation*}
	\|\Lambda(k)\|_{r_k} =  \Big(\sum_{i \in M}\Lambda_{i}^{r_k}(k)\Big)^{1/r_k} . 
\end{equation*} 
\end{definition}
\noindent
An algorithm is said to $\alpha_{r_k}$-competitive in the $L_{r_k}$ norm 
if $\|\Lambda(k)\|_{r_k} \leq \alpha_{r_k}$ 
for every dimension $k$.
Note the (necessary) difference between the definitions of \vslpi and \vslpu:
in the former, the algorithm must be competitive in all norms in all
dimensions simultaneously, whereas in \vslpu, the algorithm only 
needs to be competitive against a single norm in each dimension that
is specified in the problem input. We obtain the following result
for the any norm problem.
\begin{theorem}
\label{thm:vsany}
	There is an online algorithm for the \vslpu problem that
	simultaneously obtains a competitive ratio of $O(r_k + \log d)$
	for each dimension $k$, where the goal is to optimize the $L_{r_k}$
	norm in the $k$th dimension. Moreover, these competitive ratios
	are tight, i.e., there is a matching lower bound for every $L_r$ norm.
\end{theorem}
\subsection{Our Techniques} 

First, we outline the main techniques used for the
identical machines setting. A natural starting point for 
lower bounds is the online {\em vertex coloring} (\vc)
lower bound of Halld\'{o}rsson and Szegedy~\cite{HalldorssonS94}, 
for which connections to \vsmaxi~\cite{ChekuriK04} have previously
been exploited. The basic idea is to encode a \vc instance as a 
\vsmaxi instance where the number of dimensions $d$ is (roughly) 
$n^B$ and show that an approximation factor of (roughly) 
$B$ for \vsmaxi implies an approximation factor of 
(roughly) $n^{1-1/B}$ for \vc. 
One may want to try to combine this reduction and the online lower
bound of $\Omega(n / \log^2 n)$ for \vc  ~\cite{HalldorssonS94}  to get a better lower bound for \vsmaxi.
However, the reduction crucially relies on 
the fact that a graph with the largest clique size of at most $k$ has a chromatic number of (roughly) $O(n^{1  - 1/k})$, and this 
does not imply that the graph can be colored online with a similar number of colors.

A second approach is to explore 
the connection of \vsmaxi
with online vector bin packing 
(\vbp), where multi-dimensional items arriving 
online must be packed into a minimum number of
identical multi-dimensional bins. 
Recently, Azar {\em et al.} \cite{AzarCKS13}
obtained strong lower bounds of $\Omega(d^{1/B})$ 
where $B\geq 1$ is the capacity of each bin in every 
dimension (the items
have a maximum size of 1 on any dimension). It 
would be tempting to conjecture that the inability 
to obtain a constant approximation algorithm for the 
\vbp~ problem unless $B = \Omega(\log d)$ should yield
a lower bound of $\Omega(\log d)$ for the \vsmaxi problem.
Unfortunately, this is false. The difference between the 
two problems is in the capacity of the bins/machines
that the optimal solution is allowed to use: in \vsmaxi,
this capacity is 1 whereas in \vbp, this capacity is $B$, and 
using bins with larger capacity can decrease the number of bins needed 
super-linearly in the increased capacity. Therefore, a lower bound for \vbp does not imply any 
lower bound for \vsmaxi. On the other hand, 
an upper bound of $O(d^{1/(B-1)} \log d)$ for the 
\vbp problem is obtained in \cite{AzarCKS13}
via an $O(\log d)$-competitive algorithm for \vsmaxi. 
Improving
this ratio considerably for \vsmaxi would have been a natural approach
for closing the gap for \vbp; unfortunately, our lower
bound of $\Omega(\log d / \log  \log d)$ rules out this possibility.

Our lower bound is obtained via a different approach from the 
ones outlined above. At a high level, we leverage the connection 
with coloring, but one to a problem
of minimizing the size of the largest monochromatic clique
given a fixed set of colors. Our main technical result
is to show that this problem has a lower bound of $\Omega(\sqrt{t})$
for online algorithms, where $t$ is the number of colors. 
To the best of our knowledge, this problem was not studied before and we believe this result should be of  independent interest.\footnote{
In \cite{mutze2014coloring}, the problem of coloring vertices without creating certain monochromatic subgraphs was studied,  which is different from 
our goal of minimizing the largest monochromatic clique size. Furthermore, this previous work was only for random graphs and the focus was on whether the desirable 
coloring is achievable online depending on the parameters of the random graph. 
}
As is typical in establishing online lower bounds, the construction 
of the lower bound instance
is viewed as a game between the online algorithm and the adversary. 
Our main goal is to force the online algorithm 
to grow cliques while guaranteeing that the optimal (offline) solution can color vertices 
in a way that limits clique sizes to a constant. The technical challenge is to 
show that the optimal solution does not form large cliques across the cliques that
the algorithm has created. For this purpose, we develop a novel randomized code
that dictates the choices of the optimal solution and restricts those
of the online algorithm. Using the probabilistic method on this code,
we are able to show the existence of codewords that always lead to a good
optimal solution and an expensive algorithmic one.
We also show that the same idea can be used to obtain a lower bound for any $L_r$ norm.

\smallskip
We now turn our attention to our second main result which is in the 
unrelated machines setting: an upper bound for the 
\vslpu problem. Our algorithm is greedy with respect to 
a potential function
(as are algorithms for all special cases studied earlier
\cite{AwerbuchAGKKV95,AspnesAFPW97,Caragiannis08,MeyersonRT13}),
and the novelty lies in the choice of the potential function.
For each individual dimension $k$, we 
use the $L_{r_k}^{r_k}$ norm as the potential 
(following \cite{AspnesAFPW97,Caragiannis08}). 
The main challenge is to combine these individual potentials 
into a single potential. 
We use a weighted linear combination of the individual potentials 
for the different dimensions. This is 
somewhat counter-intuitive since the combined potential can 
possibly allow a large potential in one dimension to be compensated
by a small potential in a different one --- indeed,
 a na\"{i}ve combination only gives
a competitive ratio of $O( \max_{k} r_k + \log d)$
for all $k$. 
 However, we
observe that we are aiming for a competitive ratio of 
$O(r_k + \log d)$ which allows some slack 
compared to scalar loads if $r_k < \log d$. Suppose
$q_k = r_k + \log d$; then we use weights of
$q_k^{-q_k}$ in the linear combination 
after changing the individual potentials to $L_{r_k}^{q_k}$.
Note that as one would expect,
the weights are larger for dimensions that allow a smaller slack.
We show
that this combined potential simultaneously leads to the 
asymptotically optimal competitive ratio on every individual dimension.

\smallskip
Finally, we briefly discuss our other results. Our slightly improved upper bound for the \vsmaxi problem follows from 
a simple random assignment and redistributing `overloaded' machines. We remark that  derandomizing this strategy  
is relatively straightforward. Although
this improvement is very marginal, we feel that this is somewhat interesting 
since our algorithm is simple and perhaps more intuitive yet gives the tight upper bound. 
For the \vslpi
problem, we give a reduction to \vsmaxi by structuring the 
instance by ``smoothing'' large jobs and then arguing that for
structured instances, a \vsmaxi algorithm is also optimal
for other $L_r$ norms.

\subsection{Related Work} 

Due to the large volume of related work, we will only sample
some relevant results in online scheduling and refer the 
interested reader to more detailed 
surveys (e.g., \cite{Azar96,Sgall96,Sgall05,PruhsST04})
and textbooks (e.g., \cite{BorodinE98}). 

\medskip
\noindent 
{\bf Scalar loads.} Since the $(2-1/m)$-competitive algorithm
by Graham~\cite{Graham66} for online (scalar) load balancing
on identical machines,
a series of papers~\cite{BartalFKV95,KargerPT96,Albers99}
have led to the current best ratio of 1.9201~\cite{FleischerW00}.
On the negative side, this problem was shown to be NP-hard
in the strong sense by Faigle~{\em et al.} \cite{FaigleKT89}
and has since been shown to have a competitive ratio of 
at least 1.880~\cite{BartalKR94,Albers99,GormleyRTW00,Rudin01}. 
For other norms,
Avidor~{\em et al.}\cite{AvidorAS01} obtained 
competitive ratios of $\sqrt{4/3}$ and 
$2 - O\left(\frac{\log r}{r}\right)$ for the $L_2$
and general $L_r$ norms respectively.

For unrelated machines, Aspnes~{\em et al.} \cite{AspnesAFPW97}
obtained a competitive ratio of $O(\log m)$ for makespan 
minimization, which is asymptotically tight~\cite{AzarNR95}. 
Scheduling for the $L_2$ norm was considered by 
\cite{ChandraW75,CodyC76},
and Awerbuch~{\em et al.} \cite{AwerbuchAGKKV95} obtained
a competitive ratio of $1+\sqrt{2}$, which
was shown to be tight~\cite{CaragiannisFKKM11}.
For general $L_r$ norms,
Awerbuch~{\em et al.} \cite{AwerbuchAGKKV95}
(and Caragiannis~\cite{Caragiannis08})
obtained a competitive ratio of $O(r)$, 
and showed that it is tight up to constants. 
Various intermediate settings such as related machines (machines
have unequal but job-independent speeds) \cite{AspnesAFPW97, BermanCK00} 
and restricted assignment (each job has a machine-independent load but 
can only be assigned to a subset of 
machines) \cite{AzarNR95, CaragiannisFKKM11,ChristodoulouMS12,SuriTZ07} 
have also been studied for the makespan and $L_r$ norms.

\medskip
\noindent 
{\bf Vector loads.} The \vsmaxi problem was introduced
by Chekuri and Khanna~\cite{ChekuriK04}, 
who gave an offline approximation of $O(\log^2 d)$ and 
observed that a random assignment has a competitive ratio of
$O\left(\frac{\log dm}{\log \log dm}\right)$. 
Azar~{\em et al.} \cite{AzarCKS13}
and Meyerson~{\em et al.} \cite{MeyersonRT13} 
improved the competitive ratio to $O(\log d)$ using 
deterministic online algorithms. 
An offline $\omega(1)$ lower bound was also proved in
\cite{ChekuriK04}, and it remains open as to what the 
exact dependence of the approximation ratio on $d$
should be. Our online lower bound asserts that a 
significantly sub-logarithmic dependence would require 
a radically different approach from all the 
known algorithms for this problem.

For unrelated machines, Meyerson~{\em et al.} \cite{MeyersonRT13}
noted that the natural extension of the algorithm of 
Aspnes~{\em et al.} \cite{AspnesAFPW97} to vector loads
has a competitive ratio of $O(\log dm)$ for makespan
minimization; in fact, for identical machines, they used 
exactly the same algorithm but gave a tighter analysis.
For the offline \vsmaxu problem, 
Harris and Srinivasan~\cite{HarrisS13} recently showed
that the dependence on $m$ is not required by giving 
a randomized $O(\log d / \log \log d)$ approximation 
algorithm. 

\section{Identical Machines}
\label{sec:identical} 

First, we consider the online vector scheduling problem for identical
machines. 
\submit{We give tight upper and lower bounds
for this problem, both for the makespan norm (Theorem~\ref{thm:vsmaxi}) 
and for arbitrary $L_r$ norms (Theorem~\ref{thm:vsall}).
In the extended abstract, we only present the lower bound for 
the makespan norm and its extension to arbitrary $L_r$ norms; 
the upper bounds appear in the full paper.
}
\full{
In this section, we obtain tight upper and lower bounds
for this problem, both for the makespan norm (Theorem~\ref{thm:vsmaxi}) 
and for arbitrary $L_r$ norms (Theorem~\ref{thm:vsall}).
}

\full{
\subsection{Lower Bounds for \vsmaxi and \vslpi} 

In this section, we will prove the lower bound in Theorem~\ref{thm:vsmaxi}, 
i.e., show that any online algorithm  
for the \vsmaxi problem can be forced to construct a schedule such that there exists a dimension 
where one machine has load $\Omega(\log d / \log \log d)$, whereas the optimal
schedule has $O(1)$ load on all dimensions of all machines.
This construction will also be extended to all $L_p$ norms (\vslpi) in order to establish 
the lower bound in Theorem~\ref{thm:vsall}. 
%
}

We give our lower bound for \vsmaxi in two parts. 
First in Section \ref{subsec:mclb}, we define a lower bound instance 
for an online graph coloring problem, which we call {\sc Monochromatic Clique}. 
Next, in Section \ref{subsec:vsmaxlb}, we show how our lower bound instance for {\sc Monochromatic Clique} 
can be encoded as an instance for \vsmaxi in order to obtain the desired  $\Omega(\log d / \log \log d)$ bound. 
\medskip

\submit{
\subsection{Lower Bound for {\sc Monochromatic Clique}}
\label{subsec:mclb}
}

\full{
\subsubsection{Lower Bound for {\sc Monochromatic Clique}}
\label{subsec:mclb}
}

The {\sc Monochromatic Clique} problem is defined as follows:

%
%
\smallskip \noindent
{\sc Monochromatic Clique}: We are given a {\em fixed} set of $t$ colors. 
The input graph is  revealed to an algorithm as an online sequence of  $n$ vertices
$v_1, \ldots, v_n$ that arrive one at a time.  When vertex $v_j$ arrives, we are given 
all edges between vertices $v_1, v_2, \ldots, v_{j-1}$ and vertex $v_j$.
The algorithm must then assign $v_j$
one of the $t$ colors before it sees the next arrival.  The objective is to minimize the size of the largest
monochromatic clique in the final coloring.  
\smallskip

The goal of the section will be to prove the following lemma, which we  will  use later in Section 
\ref{subsec:vsmaxlb} to establish our lower bound for \vsmaxi. 

\begin{theorem} 
\label{lem:mono-clique}
The competitive ratio of any online algorithm for {\sc Monochromatic Clique}
is  $\Omega(\sqrt{t})$, where $t$ is the number of available colors. 
\end{theorem} 

More specifically, for any online algorithm $A$, there is an instance on which $A$ produces a monochromatic clique of
size $\sqrt{t}$, whereas the optimal solution can color the graph such that the size of the largest
monochromatic clique is $O(1)$.

We will frame the lower bound as a game between the adversary and the online algorithm. At a high level, the 
 instance is designed as follows.  For each new arriving vertex $v$ and color $c$, the adversary connects $v$
 to every vertex in some currently existing monochromatic clique of color $c$. 
 Since we do this for every color, this ensures that regardless of the algorithm's coloring of $v$, some monochromatic clique grows by 1 in size (or the first vertex in a clique is introduced). 
 Since this growth happens for every vertex, the adversary is able to 
 quickly force the algorithm to create a monochromatic clique of size $\sqrt{t}$.
 
 The main challenge now is to ensure that the adversary can still obtain a good offline solution. Our choice for this solution will be 
 na\"{i}ve: the adversary will simply properly color the monochromatic cliques it attempted to grow in the algorithm's solution. 
 Since the game stops once the algorithm has produced a monochromatic clique of size $\sqrt{t}$,
 and there are $t$ colors, such a proper coloring of every clique is possible. The risk 
 with this approach is that a large monochromatic clique may now form in the 
 adversary's coloring from edges that cross these independently grown cliques
  (in other words, properly colored cliques in the algorithm's solution could now become monochromatic cliques for the adversary). This may seem hard 
 to avoid since each vertex is connecting to some monochromatic clique for every color. However, 
in our analysis we show that if on each step the adversary selects which cliques to grow
in a carefully defined random fashion, then with positive
probability, all properly colored cliques in the algorithm's solution that hurt the adversary's na\"{i}ve solution are of size $O(1)$.  

\smallskip

\noindent
{\bf Instance Construction:} 
We adopt the standard terminology used in online coloring problems
(see, e.g.~\cite{HalldorssonS94}).
Namely, the algorithm will place each vertex in one of $t$ {\em bins} to define its color
assignments, whereas we will use {\em colors} to refer to the color assignment in the
optimal solution (controlled by the adversary).  For each vertex arrival, the game is
defined by the following 3-step process: 

\begin{enumerate} 
\setlength{\itemsep}{0pt}
\setlength{\parskip}{0pt}
\item The adversary issues a vertex $v_j$ and defines $v_j$'s adjacencies
with vertices $v_1, \ldots, v_{j-1}$.  
\item The online algorithm places $v_j$ in one of the available $t$ bins. 
\item The adversary selects a color for the vertex.  
\end{enumerate}
\noindent 


We further divide each bin into $\sqrt{t}$ {\em slots} $1,2, \ldots, \sqrt{t}$.
These slots will only be used for the adversary's bookkeeping.  
Correspondingly, we partition the $t$ colors into $\sqrt{t}$ {\em color sets}
$C_1, \ldots, C_{\sqrt{t}}$, each of size $\sqrt{t}$. Each vertex will reside
in a slot inside the bin chosen by the algorithm, and all vertices residing in
slot $i$ across all bins will be colored by the optimal solution using a color
from $C_i$. The high-level goal of the construction will be to produce 
properly colored cliques inside each slot of every bin.

Consider the arrival of vertex $v_j$.  Inductively assume the previous vertices
$v_1, \ldots, v_{j-1}$ have been placed in the bins by the algorithm, 
and that every vertex within a bin
lies in some slot.  Further assume that all the vertices in any particular
slot of a bin form a properly colored clique.  

\begin{figure} 
\begin{center} 
\resizebox{6.8in}{!}{\includegraphics{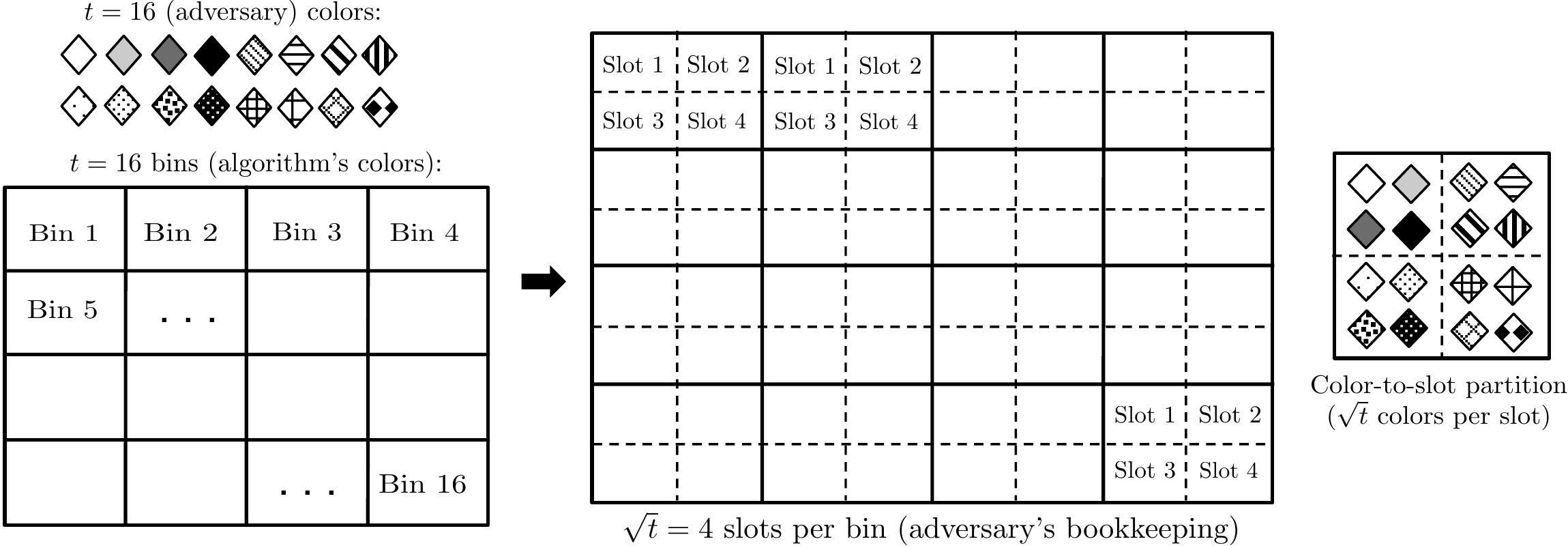}} 
\caption{{\small  Illustration of the construction set-up for Lemma \ref{lem:mono-clique}.}}
\label{fig:lb1}


\resizebox{6.8in}{!}{\includegraphics{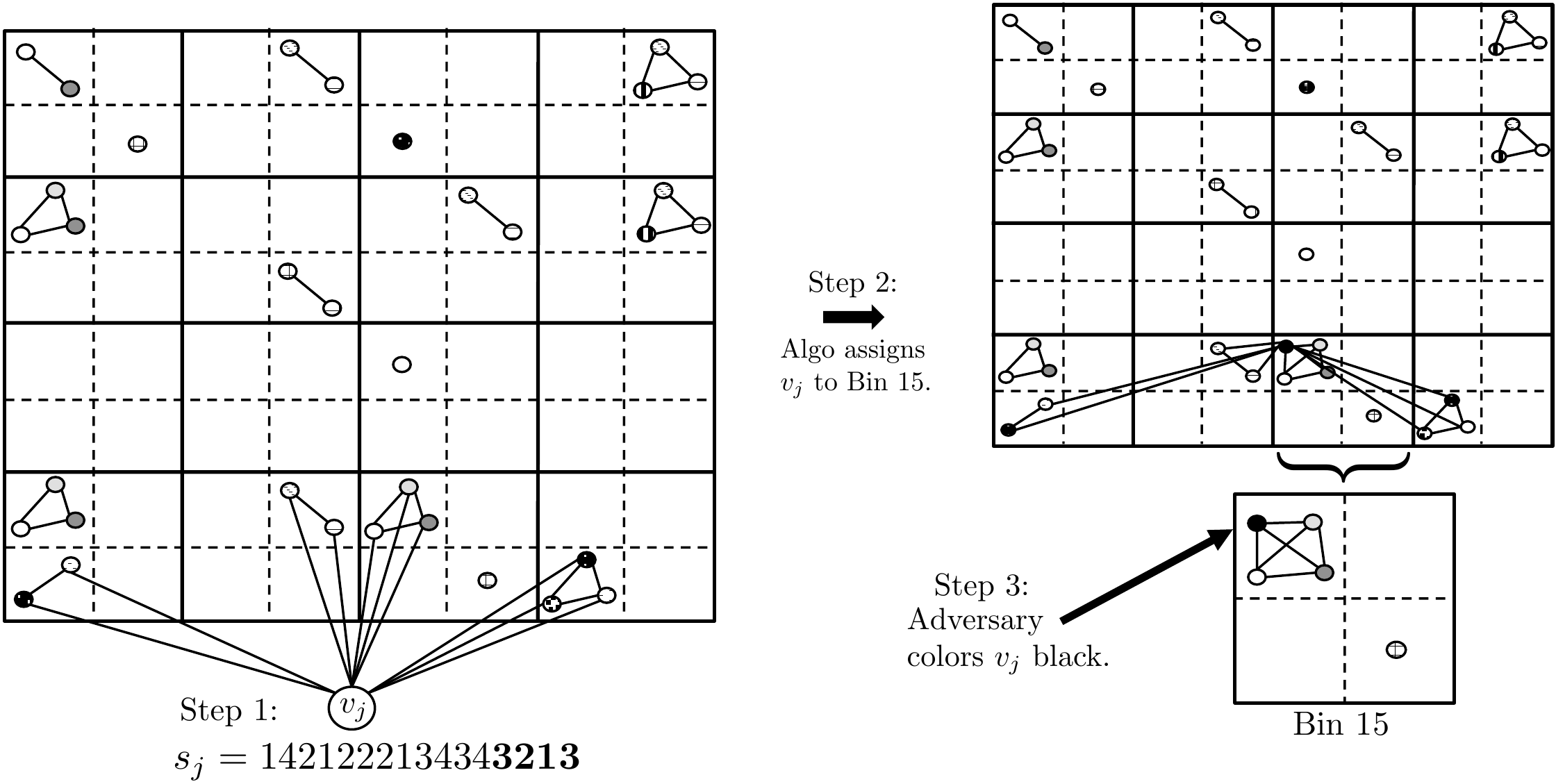}} 
{\small \caption{Depiction of the three-step lower-bound game for Lemma \ref{lem:mono-clique}. 
For simplicity, the only adjacencies shown for vertices 
issued before $v_j$ are those between vertices in the same 
bin-slot pair (in reality, other adjacencies also exist).
Also for simplicity, the only adjacencies shown for $v_j$ are those 
it has with vertices in bins 13 through 16 (dictated by the bold substring ``3213" in $s_j$).
Again note that in reality, $v_j$ will also be adjacent to vertices in bins 1 through 12 due 
to remaining prefix ``142122213434".}}\label{fig:lb2} 
\end{center}

\end{figure}

To specify the new adjacencies formed by vertex $v_j$ for Step 1, we will use a $t$-length
$\sqrt{t}$-ary string $s_j$, where we connect $v_j$ to every vertex in slot
$s_j[k]$ of bin $k$, for all $k=1, 2, \ldots, t$. 
Next, for Step 2, the algorithm places $v_j$ in some bin $b_j$. We say that $v_j$
is then placed in slot $q_j = s_j[b_j]$ in bin $b_j$.  Finally for Step 3,
the adversary chooses an arbitrary color for $v_j$ from the colors in $C_{q_j}$ 
that have not yet been used for any vertex in slot $q_j$ of bin $b_j$.
The adversary will end the instance whenever there 
exists a slot in some bin that contains $\sqrt{t}$ vertices.
This ensures that as long as the game is running, there is 
always an unused color in every slot of every bin.
Also observe that after this placement, the clique in 
slot $q_j$ in bin $b_j$ 
has grown in size by 1 but is still properly colored. 
So, this induction is well defined.  This completes description of the instance (barring our choice for each 
adjacency string $s_j$). See Figures \ref{fig:lb1} and 2 
for illustrations of the construction. 

\smallskip
\noindent{\bf Instance Analysis:} The following lemma follows directly from the construction.

\smallskip

\begin{lemma}
\label{lem:algocost}
For any online algorithm there is a monochromatic clique of size $\sqrt t$.
\end{lemma}
\begin{proof}
After $t^2$ vertices are issued,
there will some bin $b$ containing at least $t$ vertices, and therefore some 
slot in bin $b$ containing  at least
$\sqrt{t}$ vertices forming a clique of size $\sqrt{t}$.
Since all the vertices in the clique are in the same bin, there exists a monochromatic
clique of size $\sqrt t$ in the algorithm's solution.
\end{proof}

Thus, it remains  
 to show that there exists a sequence of 
$\sqrt{t}$-ary strings of length $t^2$ 
(recall that these strings define the adjacencies for 
each new vertex) such that the size of the largest 
monochromatic clique in the optimal coloring is $O(1)$. For brevity,
we call such a sequence a \emph{good sequence}.

\smallskip
\noindent 

First observe that monochromatic edges 
(i.e., edges between vertices of the same color)
cannot form between vertices in slots 
$s$ and $s'\not=s$ (in the same or in different bins) 
since the color sets used for the slots are disjoint.  
Moreover, monochromatic edges cannot form within the 
same slot in the same bin since these 
vertices always form a properly colored clique. 
Therefore, monochromatic edges can only form between 
two adjacent vertices 
$v_j$ and $v_{j'}$ such that $q_j = q_{j'}$ and 
$b_{j} \neq b_{j'}$, i.e., vertices in the same slot but in different bins.  Relating back to our earlier discussion, 
these are exactly the edges that are properly colored in the algorithm's 
solution that could potentially form monochromatic cliques in the adversary's solution; 
we will refer to such edges as {\em bad edges}. 

Thus, in order to define a good sequence of $t^2$ strings, we need ensure our adjacency strings do not induce large cliques 
of bad edges.  To do this, we first need a handle on what structure must exist across the sequence in order for bad-edge cliques to form.  
This undesired structure is characterized by the following lemma. 

\begin{lemma}
\label{l:str-clique} 
{\em Suppose $K = \{u_{\phi(1)}, \ldots, u_{\phi(w)}\}$ is a $w$-sized monochromatic clique of color $c \in C_\ell$ that 
forms during the instance, where $\phi: [w] \rightarrow [t^2]$ maps $k \in [w]$ to the index of the $k$th vertex to join $K$ 
(note, from the above discussion, that $b_{\phi(j)}$ are different for all $i \in [w]$). Then 
\begin{equation*}\label{eq:1}
s_{\phi(j)}[b_{\phi(i)}] = \ell \ \;\;\;\forall j \in \{1, \ldots , w\}, \forall i \in \{1, \ldots, j-1 \}.
\end{equation*}  }
\end{lemma}  
\begin{proof}
Consider vertex $u_{\phi(j)}$ (the $j$th vertex to join $K$).  
Since $K$ is a clique, $u_{\phi(j)}$ must be adjacent to vertices $u_{\phi(1)},
 \ldots, u_{\phi(j-1)}$.  Since all these vertices are colored with $c \in C_\ell$, they must have been placed 
in slot $\ell$ in their respective bins.  Therefore, the positions in $s_{\phi(j)}$ that correspond to these bins 
must also be $\ell$, i.e., $s_{\phi(j)}[b_{\phi(i)}] = \ell$ for all previous vertices $u_{\phi(i)}$. 
\end{proof}

In the remainder of the proof, we show that the structure in Lemma \ref{l:str-clique} can be avoided with non-zero probability for constant sized cliques if 
we generate our strings uniformly at random, thus implying the existence of a good set of $t^2$ strings.



Specifically, suppose the adversary picks each $s_j$ uniformly at random, i.e., for each character in $s_j$ we pick $w \in [\sqrt{t}]$ with probability $t^{-1/2}$.
We define the following notation: 
\begin{itemize}  
\setlength{\itemsep}{0pt}
\setlength{\parskip}{0pt}
\item  Let $K_{20}$ be the event that the adversary creates a monochromatic clique of size 20 or greater.\footnote{20 is an arbitrarily chosen large enough constant.}
\item  Let $K_{20}(S,c)$ be the event that
a monochromatic clique $K$ of color $c$ and size 20 or greater forms such that
the first 10 vertices to join $K$ are placed in the bins specified by the set of 10 indices $S$.
\item  Let $P_j(S,q)$ be a random variable that is 1 if $s_j[i] = q \ \forall i \in S$ and 0 otherwise.
 Let $P(S,q) = \sum_{j=1}^{t^2}P_j(S,q)$. 
\item Let $q(c) \in [\sqrt t]$ to be the index of the color set to which color $c$ belongs (i.e., $c \in C_{q(c)}$).  
\item Let $[n]_k := {[n] \choose k}$ denote the set of all size-$k$ subsets of $[n]$. 
\end{itemize} 

The next lemma follows from standard Chernoff-Hoeffding bounds, which we state first for completeness. 
\begin{theorem}
	\label{thm:ch} 
         (Chernoff-Hoeffding Bounds (e.g., \cite{MotwaniR97}))
	Let $X_1, X_2, ..., X_n$ be independent binary random variables and let $a_1, a_2, ..., a_n$ be coefficients in $[0, 1]$ . Let $X  = \sum_i a_i X_i$. Then, 
	\begin{itemize}
		\item For any $\mu \geq {\mathbb E}[X]$ and any $\delta >0$, $\Pr [ X > ( 1+ \delta) \mu] \leq \left( \frac{ e^\delta}{(1 + \delta)^{(1+ \delta)}}\right)^\mu$. 
		\item For any $\mu \leq {\mathbb E}[X]$ and $\delta >0$, $\Pr [X < (1 - \delta) \mu] \leq e^{-\mu \delta^2 / 2}$. 
	\end{itemize}
\end{theorem}
We are now ready to state and prove the lemma.
\begin{lemma} 
\label{lem:ilb1}
If the adversary picks each $s_j$ uniformly at random, then $\Pr[P(S,q) \geq 10] < t^{-30}$. 
\end{lemma}
\begin{proof}
First, we observe that for any set $S \in [t]_{10}$ and any $r \in [\sqrt{t}]$, 
we have $\Pr[P_i(S,r)=1] = (1/\sqrt{t})^{10} = t^{-5}$. 
Therefore by linearity of expectation, we have
\begin{alignat}{2}
{\mathbb E}[P(S,r)]  = {\mathbb E}\left[ \sum_{i=1}^{t^2}P_i(S,r)\right] 
= t^2 \cdot t^{-5} = t^{-3}. \label{eq:2} 
\end{alignat} 
Applying Theorem \ref{thm:ch} to $P(S,r)$ 
with $X_i = P_i(S, r)$, $a_i = 1$, $\delta = 10t^3 -1$ and $\mu = t^{-3}$ from  Eqn.~\eqref{eq:2}, 
we get
\begin{equation*}
\Pr[ P(S,r) \geq 10] 
\leq \left(\frac{e^{10t^3 -1}}{(10t^3)^{10t^3}}\right)^{t^{-3}}
\leq \left(\frac{e^{10}}{10^{10}}\right) \cdot \left(\frac{1}{t^{30}}\right)
< t^{-30}.\qedhere 
\end{equation*} 
\end{proof}

Using Lemmas \ref{l:str-clique} and  \ref{lem:ilb1}, we argue that there exist an 
offline solution with no monochromatic clique of super constant size. 

\begin{lemma}
\label{l:optconst}
There exists an offline solution where every monochromatic clique is of size $O(1)$.
\end{lemma}

\begin{proof}
To show the existence of a good set of $t^2$ strings,  
it is sufficient to show that $\Pr[K_{20}] < 1$. Using Lemma \ref{lem:ilb1},
we in fact show this event occurs with low probability.  Observe that

\begin{equation} 
\label{eq:k20bound} 
\Pr[K_{20}]  \leq \sum_{c \in [\sqrt{t}]} \sum_{S \in [t]_{10}} \Pr[K_{20}(S,c)]  \leq \sum_{c \in [\sqrt{t}]} \sum_{S \in [t]_{10}} \Pr[P(S,q(c)) \geq 10].
\end{equation} 
The first inequality is a straightforward union bound.  The second inequality follows by Lemma \ref{l:str-clique}.
 If the event $K_{20}(S,c)$ occurs,  then Lemma \ref{l:str-clique}  
implies $s_j[b_i] = q(c)$  for $j = 11, \ldots, 20$, $i \in S$.  

Since there are $\sqrt{t}$ possible colors and $|[t]_{10}| < t^{10}$,
applying both \eqref{eq:k20bound}  and Lemma \ref{lem:ilb1} we get
\begin{equation*} 
\Pr[K_{20}] \leq \sum_{c \in [\sqrt{t}]} \sum_{S \in [t]_{10}} \Pr[P(S,q(c)) \geq 10]  \leq \sum_{c \in [\sqrt{t}]} \sum_{S \in [t]_{10}} t^{-30}
\leq  t^{1/2}\cdot t^{10} \cdot t^{-30} = t^{-39/2}  < 1, \label{eq:43}
\end{equation*}
for all $t > 1$. Therefore, there is an optimal coloring such that there is no monochromatic clique of size more than 20.
\end{proof}

Theorem~\ref{lem:mono-clique} now follows directly from 
Lemmas~\ref{lem:algocost} and \ref{l:optconst}.

\medskip
\submit{
\subsection{Lower Bound for \vsmaxi and  \vslpi from {\sc Monochromatic Clique}}
\label{subsec:vsmaxlb} 
}

\full{
\subsubsection{Lower Bound for \vsmaxi and  \vslpi from {\sc Monochromatic Clique}}
\label{subsec:vsmaxlb} 
}

We are now ready to use Theorem \ref{lem:mono-clique} to show an
$\Omega(\log d/ \log \log d)$ lower bound for \vsmaxi.
 We will describe a lower bound instance for \vsmaxi whose structure is
based on an instance of {\sc monochromatic clique}. This will allow us to use the lower bound 
instance from Theorem \ref{lem:mono-clique} as a black box to produce the desired lower bound
for \vsmaxi. 

\begin{figure} 
\begin{center} 
\resizebox{6.7in}{!}{\includegraphics{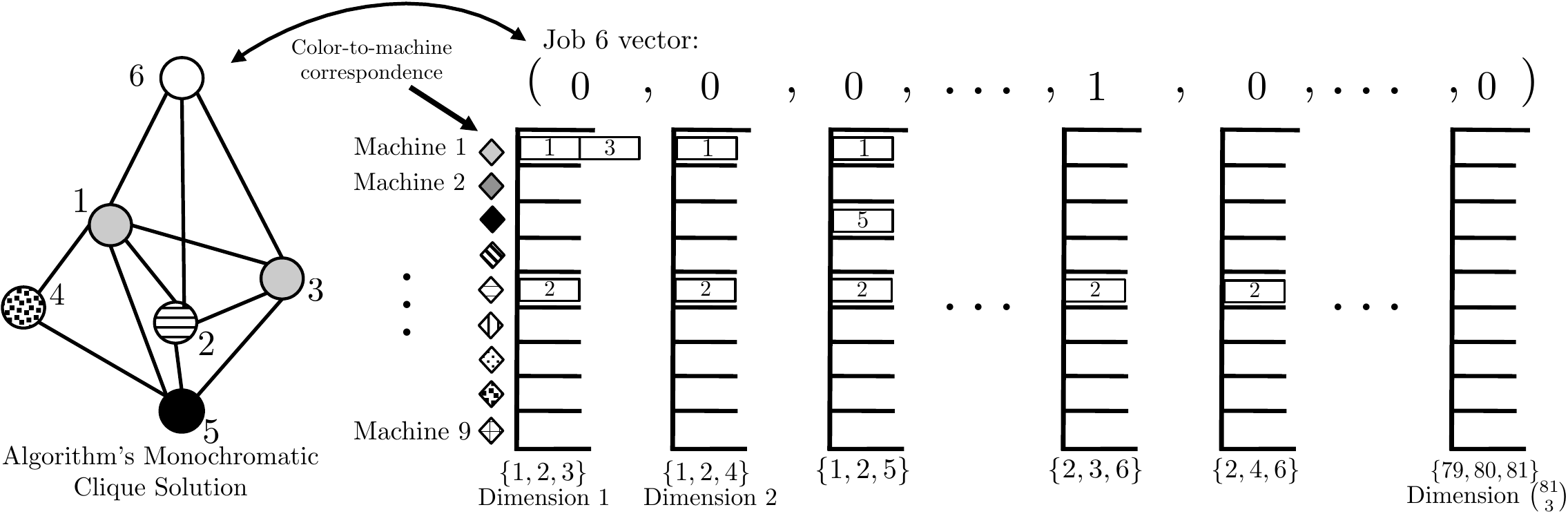}} 
\caption{{\small Illustration of the lower bound construction for \vsmaxi using the {\sc monochromatic clique} lower bound (Theorem \ref{lem:mono-clique}) for an instance where 
$m=9$ (and thus $d= {9^2 \choose \sqrt{9} }= {81 \choose 3}$ for the \vsmaxi instance and $t = 9$ for the {\sc monochromatic clique} instance).  Currently job 6 is being issued; its binary load 
vector, which is based on the current edge structure in the {\sc monochromatic clique} instance, is given above the machines/dimensions.  Observe that job 6 has 0 load in the first three dimensions and the last dimension since 6 is not contained in any of the these dimensions' $S_k$ sets (indicated below). It does have load 1 in the dimension corresponding to set $\{2,3,6\}$ since vertex 6 forms a clique with vertices 2 and 3 in the {\sc monochromatic clique} instance; however, it still has load 0 in dimension $\{2,4,6\}$ since vertex 6 does not form a clique with vertices 2 and 4. }}
\label{fig:lb3} 


\resizebox{6.7in}{!}{\includegraphics{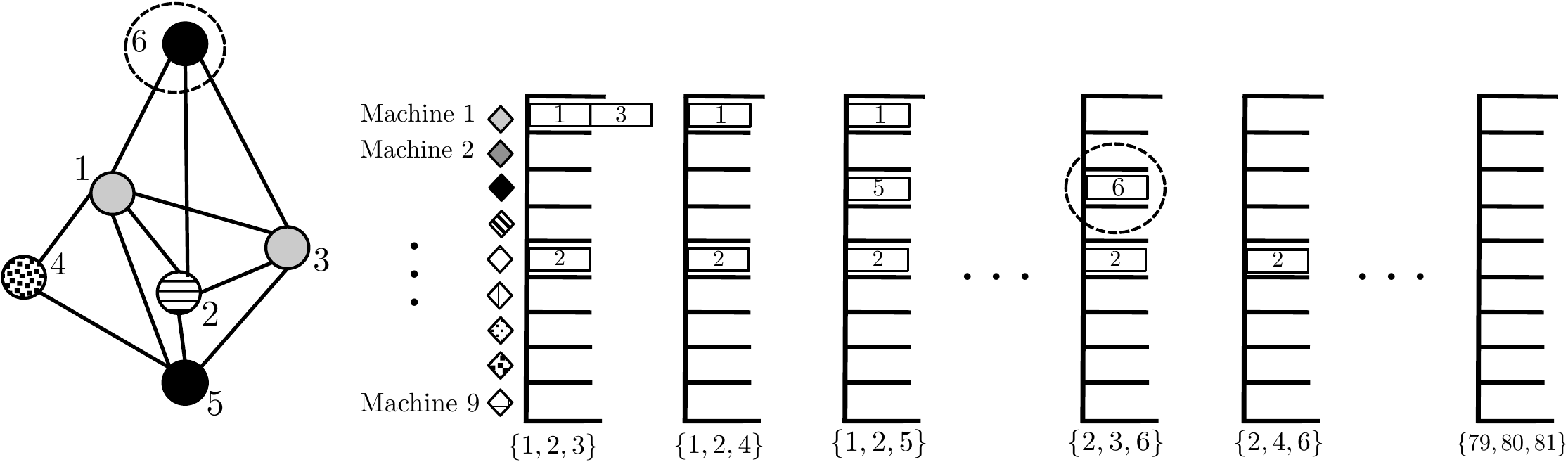}} 
\caption{{\small State of the construction after job 6 is assigned to machine 3. Since black is the color we associated with machine 3, this job assignment by 
the \vsmaxi algorithm is translated as coloring vertex 6 black in the {\sc monochromatic clique} instance.}}
\label{fig:lb4}

\end{center}

\end{figure}

We first set the problem definition of {\sc Monochromatic clique} to be for $m$ colors where $m$ is also the number of machines used in the \vsmaxi instance. 
Let $I_C$ be the lower-bound instance for this problem given by Theorem \ref{lem:mono-clique}. This 
produces a graph $G$ of $m^2$ vertices such that the algorithm forms a monochromatic clique of size $\sqrt{m}$, 
whereas the largest monochromatic clique in the optimal solution is of size $O(1)$.  Let $G_j = (V_j, E_j)$ be the graph 
in $I_C$ after vertices $v_{1}, \ldots, v_{j}$ have been issued (and so $G_n = G$).
We define the corresponding lower bound instance for \vsmaxi as follows (see Figures 3 
and \ref{fig:lb4} for an illustration): 

\begin{itemize} 
\setlength{\itemsep}{0pt}
\setlength{\parskip}{0pt}
\item There are $m^2$ jobs, which correspond to vertices $v_1, \ldots, v_{m^2}$ from $I_C$.
\item Each job has $d = \binom{m^2}{\sqrt{m}}$ dimensions, where each dimension 
corresponds to a specific $\sqrt{m}$-sized vertex subset of the $m^2$ vertices. Let 
$S_1, \ldots, S_d$  be an arbitrary ordering of these subsets. 
\item Job vectors will be binary. Namely, the $k$th vector entry for job $j$ 
is 1 if $v_j \in S_k$ and the vertices in $\{v_1, \cdots, v_j\}  \cap S_k$ form a clique in $G_j$ (if $\{v_1, \cdots, v_j\} \cap S_k = \{v_j\}$, then it is considered a 1-clique); 
otherwise,  the $k$th entry is 0. 
\item Let $c_1, \ldots, c_m$ define an ordering on the available colors from $I_C$. 
We match each color from $I_C$ to a machine in our scheduling instance. 
Therefore, when the \vsmaxi  algorithm makes an assignment for a job, we translate this machine assignment as the corresponding color assignment in $I_C$. 
Formally, if job $j$ is placed on machine $i$ in the scheduling instance, then vertex $v_j$ is assigned color $c_i$ in $I_C$. 
\end{itemize}

\smallskip
\noindent
Since assigning jobs to machines corresponds to colorings in $I_C$, it follows 
that the largest load in dimension $k$ is the size of the largest monochromatic sub-clique in $S_k$.  
$I_C$ is given by the construction in Theorem \ref{lem:mono-clique}; therefore
at the end of the instance, there will exist a dimension $k'$ such that the 
online algorithm colored every vertex in $S_{k'}$ with some color $c_i$. Thus, 
machine $i$ will have $\sqrt{m}$ load in dimension $k'$. In contrast, Theorem \ref{lem:mono-clique} ensures that all the monochromatic cliques in the optimal solution 
are of size $O(1)$, and therefore the load on every machine in dimension $k'$
is $O(1)$.

 

The relationship between $m$ and $d$ is given as follows.
\begin{fact} 
\label{lem:d-m-asym}
If $d = \binom{m^2}{\sqrt{m}}$, then $\sqrt{m} = \Omega(\log d/ \log \log d)$. 
\end{fact} 

\begin{proof} 
We will use the following well-known bounds on ${n \choose k }$:
for integers $0 \leq k \leq n$,
$\left( \frac{n}{k} \right)^{k} \leq \binom{n}{k} \leq \left( \frac{en}{k} \right)^{k}$. 
First, we observe that
\begin{equation} 
\log d 
= \log\binom{m^2}{\sqrt{m}}  
\leq \log \left(\frac{em^{2}}{\sqrt{m}}\right)^{\sqrt{m}}  
= \log(e^{\sqrt{m}} \cdot m^{(3/2)\sqrt{m}}) 
= \sqrt{m}\cdot(1 + (3/2)\log m). \label{eq:5} 
\end{equation} 
We also have
\begin{equation} 
\log \log d 
= \log \log \binom{m^2}{\sqrt{m}} 
\geq \log \log \left(\frac{m^2}{\sqrt{m}}\right)^{\sqrt m} 
\geq  \log((3/2) \sqrt{m}\log m) 
\geq (1/2) \log m. \label{eq:6} 
\end{equation} 
Hence, combining Eqns.~(\ref{eq:5}) and (\ref{eq:6}), we obtain
\begin{equation*} 
\sqrt{m} 
\geq \frac{\log d}{1 + (3/2) \log m}           
\geq \frac{\log d}{1 + 3 \log \log d},
\end{equation*}
which implies that $\sqrt{m} = \Omega(\log d/ \log \log d)$, as desired.
\end{proof}



To end the section, we show that our lower bound for \vsmaxi extends 
to general $L_r$ norms (Theorem \ref{thm:vsall}). As before, our lower bound construction 
forces any algorithm  to schedule jobs so 
that there exists a dimension $k'$ where at least one machine has load 
at least $\sqrt{m}$, whereas the load on every dimension of every machine in the optimal solution 
is bounded by some constant $C$.  Since any dimension 
has at most $\sqrt{m}$ jobs with load 1, any assignment
ensures that there are at most $\sqrt{m}$ machines with non-zero load in a  given
dimension. Therefore, in the optimal solution, 
the $L_{r}$-norm of the load vector for dimension $k'$ is at most
$(C^{r} \cdot \sqrt{m})^{1/r} = C \cdot m^{1/(2r)}$. 
 
Thus, the ratio between the objective of the solution produced by the 
online algorithm and the optimal solution is at least
$m^{1/2}/(C \cdot m^{1/2r}) =(1/C)\cdot m^{(r-1)/(2r)}$.
Using Fact \ref{lem:d-m-asym}, we conclude the lower bound.


\full{
\subsection{Upper Bounds for \vsmaxi and \vslpi} 

In this section we prove the upper bounds in Theorem \ref{thm:vsmaxi} (\vsmaxi) and Theorem \ref{thm:vsall} (\vslpi). 
First, we give a randomized $O(\log d / \log \log d)$-competitive online algorithm for \vsmaxi (Section \ref{subsec:vsmaxi-rand}) and then show how to derandomize it (Section \ref{subsec:vsmaxi-derand}).
Next, we give an $O( ( \log d /\log \log d)^{\frac{r-1}{r}})$-competitive algorithm for \vslpi (Section \ref{subsec:vslpi}), i.e., 
for each dimension $k$ and $1 \leq r \leq \log m$, $\| \Lambda(k)\|_r$ is competitive with the optimal schedule for dimension $k$ under the $L_r$ norm objective.


Throughout the section 
we assume that a priori, the online algorithm is aware of both the final volume of all jobs on each dimension and the largest load over all dimensions and jobs. 
We note that the lower bounds claimed in Theorems~\ref{thm:vsmaxi} and \ref{thm:vsall} are robust against this assumption since the optimal makespan is always a constant and this knowledge does not help the online algorithm. Furthermore, these assumptions can be completely removed for our \vsmaxi algorithm by updating a threshold on the maximum job load on any dimension and the total volume of jobs that the algorithm has observed so far. 
However, in order to make our presentation more transparent and our notation simple, we present our results under these assumptions. 

For each job $j$ that arrives online, both our \vsmaxi and \vslpi algorithms will perform the following transformation: 

\smallskip
{\bf $\bullet$ Transformation 1:} Let $V = \langle V_1, \ldots, V_d\rangle$ be the volume vector given to the algorithm a priori, where 
$V_k$ denotes the total volume of all jobs for dimension $k$.  For this transformation, we normalize $p_j(k)$ by dividing it 
by $V_k/m$ (for ease of notation, we will still refer to this normalized 
value as $p_j(k)$).  
\smallskip

Our \vsmaxi and \vslpi algorithms will also perform subsequent transformations; however, these transformations will differ slightly for the two algorithms. 

\medskip 

\subsubsection{Randomized Algorithm for \vslpi}
\label{subsec:vsmaxi-rand} 

\newcommand{\cJ}{\mathcal{J}}


We now present our randomized $O(\log d / \log \log d)$-competitive algorithm for \vsmaxi. Informally, our algorithm works as follows. For each job $j$, 
we first attempt to assign it to a machine $i$ chosen uniformly at random; however, if the resulting assignment would result in a load larger than $O(\log/ \log \log d)$ on machine $i$,
then we dismiss the assignment and instead assign $j$ greedily among other previously dismissed jobs. In general, a greedy assignment can be as bad as $\Omega(d)$-competitive; however, 
in our analysis we show that a job is dismissed by the random assignment with low probability. Therefore in expectation, the total volume of these jobs is low enough to assign greedily and 
still remain competitive. 

\smallskip
\noindent {\bf Instance Transformations:} Before formally defining our algorithm, we define additional online transformations and outline
the properties that these transformations guarantee.  Note that we perform these transformations for both the randomized algorithm presented in this section and the derandomized algorithm 
presented in Section \ref{subsec:vsmaxi-derand}. These additional transformations are defined as follows (which are preformed in sequence after Transformation 1):

\begin{itemize}
\item {\bf Transformation 2}: Let $T$ be the load of the largest job in the instance (given a priori).  If for dimension $k$ we have $T \geq V_k/m$, then for each job $j$ we set $p_j(k)$
to be $(p_j(k)\cdot V_k) /(mT)$. In other words, we normalize jobs in dimension $k$ by $T$ instead of $m/V_k$. 
\item {\bf Transformation 3:} For each job $j$ and dimension $k$, if $p_j(k) < (1/d) \max_{k'} p_j(k')$, then we increase $p_j(k)$ to $(1/d) \max_{k'} p_j(k')$. 
\end{itemize}
\smallskip
Observe that after we apply Transformations 1 and 2 to all jobs, we have 
 $\sum_{j } p_j(k) \leq m$ for all $k \in [d]$ and $0 \leq p_j(k) \leq 1$ for all jobs $j$ and $k \in [d]$.

In Lemmas \ref{lem:trans-3} and \ref{lem:trans-3-property-1}, we prove additional properties that Transformation 3 preserves. 
Since Transformations 1 and 2 are simple scaling procedures, an $\alpha$-competitive algorithm on the resulting scaled instance is also $\alpha$-competitive 
on the original instance, if we only apply the first two transformations.  In Lemma \ref{lem:trans-3}, we prove that this property is still maintained after Transformation 3. 

\begin{lemma}
\label{lem:trans-3}
After Transformations 1 and 2 have been applied, Transformation 3 increases the optimal makespan by a factor of at most 2. 
\end{lemma}
\begin{proof}
Fix a machine $i$ and a dimension $k$.  Let {\sc OPT} denote the optimal assignment before Transformation 3 is applied. Let $J^*(i)$ denote the jobs assigned to machine $i$ in {\sc OPT}, 
$\Lambda^*_i(k)$ be the load of {\sc OPT} on machine $i$ in dimension $k$, 
and $\Lambda^* = \max_{i,k} \Lambda^*_i(k)$ denote the makespan of {\sc OPT}.  We will show that Transformation 3 can increase the load on machine $i$ in dimension $k$ by at most $\Lambda^*$. 

Let $V_{i}^* = \sum_{j \in J^*(i)} \sum_{k' \in [d]} p_j(k')$ denote the total volume of jobs that {\sc OPT} assigns to machine $i$.  Observe that by 
a simple averaging argument, we have $V_{i}/d \leq  \max_{k' \in [d]} \Lambda^*_i(k')$. Since Transformation 3 can increase the load of a job $j$ in a fixed dimension by at most $(1/d)\max_{k'} p_{j}(k')$, we can upper bound the total increase in load on machine $i$ in dimension $k$ as follows:

\begin{equation}
\sum_{j \in J^*(i)}(1/d)\max_{k'} p_{j'}(k') \leq V_i^*/d \leq \ \max_{k' \in [d]} \Lambda^*_i(k') \leq \Lambda^*, 
\end{equation}  
as desired. Note that the first inequality follows from the fact that the sum of maximum loads on a machine is at most the total volume of its jobs. 
\end{proof}


Recall that after Transformations 1 and 2,  $\sum_{j} p_j(k) \leq m$ for all $k \in [d]$. In Lemma \ref{lem:trans-3-property-1}, we show 
that this property is preserved within a constant factor after Transformation 3. 

\begin{lemma}
\label{lem:trans-3-property-1}
After performing Transformation 3, $\sum_{j} p_j(k) \leq 2m$ for all $k \in [d]$.
\end{lemma}
\begin{proof}
Consider any fixed dimension $k \in [d]$. After Transformation 3, each job $j$'s load on dimension $k$ increases by at most $(1 / d) \max_{k'} p_j(k')$. Hence 
the total increase in load from jobs in dimension $k$ is at most 
\begin{equation*}
\sum_j (1 / d) \max_{k'} p_j(k') \leq (1 / d) \sum_j \sum_{k' \in [d]} p_j(k') \leq (1 / d) md \leq m,
\end{equation*}  
where the second inequality and the lemma follow from the fact that $\sum_{j} p_j(k) \leq m$ before Transformation 3. 
\end{proof}

In summary, the properties that we collectively obtain from these transformations are as follows:
\begin{itemize}
\item \textbf{Property 1}. For all $k \in [d]$, $\sum_{j} p_j(k) \leq 2m$. 
\item \textbf{Property 2}. For all $j$ and $k \in [d]$, $0 \leq p_j(k) \leq 1$. 
\item \textbf{Property 3}. For all $j$ and $k \in [d]$, $(1/d) \max_{k'} p_j(k') \leq p_j(k) \leq \max_{k'} p_j(k')$. 
\item \textbf{Property 4}. The optimal makespan is at least 1. 
\end{itemize}

Property 1 is a restatement of Lemma~\ref{lem:trans-3-property-1}. Property 2 was true after the first two transformations, and Transformation 3 has no effect on this property. Property 3 is a direct consequence of Transformation 3. 

To see why Property 4 is true, let $j$ be the job with the largest load $T$ in the instance, and let $k = \arg\max_{k'} p_j(k')$ 
(i.e., $\max_{k'} p_j(k') = T$). If Transformation 2 is applied to dimension $k$, then $p_j(k) = 1$ afterwards, which immediately implies Property 4.
Otherwise, only Transformations 1 and 3 are applied to dimension $k$ and we have $\sum_{j' }p_{j'}(k) \geq m$, which
again leads to Property 4 by a simple volume argument. Thus, by Property 4 and Lemma \ref{lem:trans-3},
it sufficient to show that the makespan of the algorithm's schedule is $O(\log d/\log \log d)$. 

\smallskip 
\noindent {\bf Algorithm Definition:} 
As discussed earlier, our algorithm consists of two procedures: a random assignment and greedy packing.  It will be convenient to assume that the algorithm has two disjoint sets $M_1$, $M_2$ of  $m$ identical machines that will be used independently by the two procedures, respectively. Each machine in $M_1$ is paired with an arbitrary distinct machine in $M_2$, and the actual load on a machine will be evaluated as the sum of the loads on the corresponding pair of machines. In other words, to show competitiveness it is sufficient to prove that all machines in both $M_1$ and $M_2$ have load $O(\log d / \log \log d)$.

Define the parameter $\alpha := \frac{10 \log d}{\log  \log d}$. Our two procedures are formally defined as follows.

\begin{itemize}
	\item \textbf{First procedure (random assignment)}: Assign each job to one of the machines in $M_1$ uniformly at random. Let $J_j^1(i)$ denote the subset of the first $j$ jobs 
	$\{1, 2, ..., j\}$ that are assigned to machine $i$ in this procedure, and let $\Lambda^1_{i, j}(k)$ denote the resulting load on machine $i$ on dimension $k$ due to jobs in $J_j^1(i)$. If $\Lambda^1_{i, j}(k) \geq 2\alpha +1$ for some $k \in [d]$, then we pass job $j$ to the second procedure. (However, note that {\em all} jobs are still scheduled by the first procedure; so even if a job $j$ is passed to the second procedure after being assigned to machine $i$ in the first procedure, $j$ still contributes load to $\Lambda_i^1(k)$ for all $k$). 


	\item \textbf{Second procedure (greedy packing)}: This procedure is only concerned with the jobs $J^2$ that are passed from the first procedure.  It allocates each job in $J^2$ (in the order
	that the jobs arrive in) to one of the machines in $M_2$ such that the resulting makespan, $\max_{i \in M_2, k \in [d]} \Lambda^2_{i,j}(k)$  is minimized; $\Lambda^2_{i,j}(k)$ is analogously defined for this second procedure as above. 
\end{itemize}

This completes the description of the algorithm. We will let $J^1(i) := J^1_n(i)$ and $\Lambda^1_{i}(k) := \Lambda^1_{i, n}(k)$, and define $J^2(i)$ and $\Lambda^2_{i}(k)$ similarly. 
We emphasize again that jobs in $J^2$ are scheduled only on machines $M_2$; all other jobs are scheduled on $M_1$ machines.

\smallskip
\noindent {\bf Algorithm Analysis:} It follows directly from the definition of the algorithm that the loads on machines in $M_1$ are at most $2\alpha + 1 = O(\log d/ \log \log d)$. 
Therefore, we are only left with bounding the loads on machines in $M_2$. The following lemma shows that the second procedure receives only a small fraction of the total volume, 
which then allows us to argue that the greedy assignment in the second procedure is $\alpha$-competitive.  
\begin{lemma}
\label{lem:vol-reduction}
The probability that a job $j$ is passed to the second procedure is at most $1 / d^3$, i.e. $\Pr [ j \in J^2 ] \leq 1 / d^3$. 
\end{lemma}
\begin{proof}
Fix a machine $i$, job $j$ and dimension $k$.  Suppose job $j$ was assigned to machine $i$ by the first procedure and is passed 
to the second procedure because we would have had $\Lambda^1_{i, j}(k) \geq 2\alpha +1$. 
Since $p_j(k) \leq$ 1 due to Property 2, it follows that $\Lambda^1_{i,j-1}(k) \geq 2 \alpha$.  Therefore we will show
\begin{equation}
	\label{eq:39}
	\Pr [ \Lambda^1_{i,j-1}(k) \geq 2 \alpha] \leq 1 / d^4,
\end{equation}
where the probability space is over the random choices of jobs $1, 2, ..., j-1$. Once inequality \eqref{eq:39} is established, the lemma follows from a simple union bound over all dimensions. 

To show (\ref{eq:39}), we use standard Chernoff-Hoeffding bounds (stated in Theorem~\ref{thm:ch} earlier). Note that ${\mathbb E}[\Lambda^1_{i,j-1}(k)] \leq 2$ due to Property 1 and the fact that jobs are assigned to 
machines uniformly at random. To apply the inequality, we define random variables $X_1, X_2, ..., X_{j-1}$ where $X_{j'} =1$ if job $j'$ is assigned to machine $i$; otherwise $X_{j'} = 0$. Set the 
parameters of Theorem~\ref{thm:ch} as follows: $a_{j'} = p_{j'}(k)$, $\mu = 2$, and $\delta  = \alpha - 1$. Thus we have: 
\begin{align*}
	\Pr [ \Lambda^1_{i,j-1}(k) \geq 2 \alpha] 
	&= \Pr \left[ \sum_{j' \in [j-1]} a_{j'} X_{j'} \geq \alpha \mu \right] = \Pr \left[ \sum_{j' \in [j-1]} a_{j'} X_{j'} \geq (1 + \delta) \mu \right] \\
	&\leq \left( \frac{ e^\delta}{(1 + \delta)^{(1+ \delta)}}\right)^\mu  \leq   \frac{ e^\delta}{(1 + \delta)^{(1+ \delta)}}  \\
	&\leq 1 / (5\log d / \log \log d)^{(5\log d / \log \log d)} \leq 1 / d^4  \hspace{4mm} \text{(for sufficiently large $d$),}
\end{align*}
as desired. 
\end{proof}

Next, we upper bound the makespan of the second procedure in terms of its total volume of jobs $V_2$,
i.e. $V_2 = \sum_{j \in J^2, k \in [d]} p_j(k)$.
\begin{lemma}
	\label{lem:vol-effect}
	$\max_{i \in M_2, k \in [d]}  \Lambda^2_i(k)  \leq V_2/m + 1$.
\end{lemma}
\begin{proof}
For sake of contradiction, suppose that  at the end of the instance there exists a dimension $k$ and machine $i$ such that $\Lambda^2_i(k)  >  V_2/m  + 1$.  Let $j$ 
be the job that made machine $i$ first cross this $V_2/m + 1$ threshold in dimension $k$.   For each machine $i'$, let $k_{i'} = \arg\max_{k'} \Lambda_{i,j-1}^2(k')$ denote the dimension with maximum load on machine $i'$ before $j$ was assigned. 


By Property 2 and the greediness of the algorithm, we have that $\Lambda_{i',j-1}^2(k_{i'}) > V_2/m$ for all $i'$. Otherwise, $j$ would have been assigned to a
machine other than $i$ resulting in a makespan less than $V_2/m  + 1$ (since $\max_{k,j} p_j(k)  \leq 1$). However, this implies that every machine in $M_2$ has a dimension with more than $V_2/m$
load. Clearly, this contradicts the definition of $V_2$. 
\end{proof}

We are now ready to complete the analysis. From Lemma~\ref{lem:vol-reduction} and linearity of expectation, we know that
\begin{equation} 
\label{eq:expmakespan}
{\mathbb E} [V_2] \leq \frac{1}{d^3} \sum_{j, k \in [d]} p_j(k) \leq \frac{1}{d^3} \cdot 2dm = \frac{m}{d^2}, 
\end{equation} 
where the second inequality follows from Property 1. Hence, inequality \eqref{eq:expmakespan} along Lemma~\ref{lem:vol-effect} imply 
that the second procedure yields an expected makespan of $O(1)$, which completes our analysis. 

\medskip

\subsubsection{Derandomized Algorithm for \vsmaxi}
\label{subsec:vsmaxi-derand}

Our derandomization borrows the technique developed in \cite{BuchbinderN05}. To derandomize the algorithm, we replace the first procedure --- a uniformly random assignment --- with a deterministic assignment guided by the following potential $\Phi$. Let $f(x) := \alpha^x$ for notational simplicity. Recall that  $\alpha := 10 \log d / \log \log d$. 
\begin{align*}
	\Phi_{i,k}(j) &:= f\left( \Lambda^1_{i,j}(k) - \frac{\alpha}{m} \sum_{j' \in [j]} p_{j'}(k)\right) &&\forall i \in M_1, j\in [n], k \in [d] \\
	\Phi(j) &:=  \sum_{i \in M} \sum_{k = 1}^d \Phi_{i,k}(j)
\end{align*}

\begin{itemize}
	\item \textbf{(New deterministic) first procedure.} Each job $j$ is assigned to a machine $i$ such that $\Phi(j)$ is minimized. If $\Lambda_{i,j}(k) \geq 3\alpha +1$, then $j$ is added to queue $J^2$ so that it can be scheduled by the second procedure. As before, each job is scheduled by either the first procedure or the second, and  contributes to the ``virtual" load $\Lambda_{i,j}(k)$ in either case. 
\end{itemize}

\begin{lemma}
	\label{lem:potential-monotone}
	$\Phi(j)$ is non-increasing in $j$. 
\end{lemma}
\begin{proof}
	Consider the arrival of job $j$. To structure our argument, we assume the algorithm still assigns $j$ to a machine in $M_1$ uniformly at random. Our goal now is to show that ${\mathbb E}[\Phi(j)] \leq \Phi(j-1)$, which implies the existence of a machine $i$ such that assigning job $j$ to the machine $i$ leads to $\Phi(j) \leq \Phi(j-1)$ (and such an assignment is actually found by the algorithm since its assignment maximizes the decrease in potential). We bound ${\mathbb E}[ \Phi_{i,k}(j)]$ as follows. 
\begin{align}
	E [ \Phi_{i,k}(j)] 
	& = \; \frac{1}{m} f \left( \Lambda_{i, j-1} + p_j(k) - \frac{\alpha}{m} p_j(k) - \frac{\alpha}{m} \sum_{j' \in [j-1]} \Lambda_{i,j-1}(k)\right) \notag\\
	& \;\;\;+  (1 - \frac{1}{m}) f \left( \Lambda_{i, j-1}  - \frac{\alpha}{m} p_j(k) - \frac{\alpha}{m} \sum_{j' \in [j-1]} \Lambda_{i,j-1}(k)\right) \notag\\
	& = \; \Phi_{i,k}(j-1) \cdot \alpha^{ - \frac{\alpha}{m} p_j(k)} \cdot \left(\frac{1}{m} (\alpha^{p_j(k)}  -1 )+ 1\right) \notag \\
	& \leq \; \Phi_{i,k}(j-1) \cdot \alpha^{ - \frac{\alpha}{m} p_j(k)} \left(\frac{p_j(k)}{m} (\alpha - 1)+ 1 \right) \label{50-1}\\
	& \leq \; \Phi_{i,k}(j-1) \cdot \exp \left( - (\alpha \log \alpha) \cdot \frac{p_j(k)}{m} \right) \exp \left(\frac{p_j(k)}{m} \cdot (\alpha - 1)\right) \label{50-2}\\
	& \leq \; \Phi_{i,k}(j-1)  \notag 
\end{align}
 Inequality (\ref{50-1}) follows since $\alpha^x -1 \leq (\alpha -1) x$ for $x \in [0, 1]$, and $p_j(k) \leq 1$ due to Property 2.  Inequality (\ref{50-2}) follows from the fact
 that $x+1 \leq e^x$. Therefore,  by linearity of expectation, we have ${\mathbb E}[\Phi(j)] \leq \Phi(j-1)$, thereby proving the lemma. 
\end{proof}

The next corollary follows from Lemma~\ref{lem:potential-monotone} and the simple observation that $\Phi(0) = md$.

\begin{corollary}
	\label{coro:phi-end}
	$\Phi(n) \leq md$. 
\end{corollary}

As in Section \ref{subsec:vsmaxi-rand}, it is straightforward to see that the algorithm forces machines in $M_1$ to have makespan $O(\alpha)$, so we
again focus on the second procedure of the algorithm. Here, we need a deterministic bound on the total volume $V_2 =  \sum_{j \in J^2} \sum_{k \in [d]} p_j(k)$ that can be scheduled on 
machines in $M_2$.  Lemma \ref{lem:volbound} provides us with such a bound. 

\begin{lemma}
\label{lem:volbound}
	$V_2 \leq m / d$. 
\end{lemma}
\begin{proof}
Consider a job $j \in J_2$ that was assigned to machine $i$ in the first procedure.  Let $k(j)$ be an arbitrary dimension $k$ 
with $\Lambda^1_{i,j} \geq 3\alpha +1$ (such a dimension exists since $j \in J^2$). 
Let $J^2_i(k) = \{j: j \in J^1(i) \cap J^2 \text{ and } k(j) = k\}$ denote the
set of jobs $j \in J^2$ that were assigned to machine $i$ by the first procedure and are associated with dimension $k$. 
We upper bound $V_2$ as follows: 
\begin{alignat}{2} 
V_2 &= \; \sum_{j \in J^2}\sum_{k' \in [d]} p_j(k')  \notag \\
&= \; \sum_{i \in M_1, k \in [d]} \sum_{ j \in J^2_i(k)} \sum_{k' \in [d]} p_j(k') && \hspace{5mm} \text{(since we associate job $j \in J^2$ with a unique dimemsion \ $k(j)$)} \notag\\
&\leq \;\sum_{i \in M_1, k \in [d]} \sum_{ j \in J^2_i(k)}  d^2 p_j(k)  && \hspace{5mm} \text{(by Property 3)}  \notag \\
&\leq \; d^2 \sum_{i \in M_1} \sum_{k \in [d]} (\Lambda^1_i (k) - 3\alpha)_+ \label{51-3} 
\end{alignat}

To see why the last inequality holds, recall that $\Lambda^1_{i,j}(k) \geq 3\alpha +1$ when $j \in J^1(i)$ and $k = k(j)$. This can happen only when $\Lambda^1_{i,j-1}(k) \geq 3\alpha$ since $p_j(k) \leq 1$ due to Property 2. Since $\Lambda^1_{i,j'}(k)$ is non-decreasing in $j'$, the sum of $p_j(k)$ over all such jobs $j$ is at most $(\Lambda^1_i (k) - 3\alpha)_+$; here $(x)_+ := \max\{0, x\}$.

We claim that for all $i \in M_1, k \in [d]$, 
\begin{align}
	\label{52}
	\Phi_{i,k}(n)  \geq \alpha^\alpha (\Lambda^1_{i,j}(k) - 3\alpha)_+
\end{align}
If $\Lambda^1_{i,j}(k) - 3\alpha \leq 0$, then the claim is obviosuly true since $\Phi_{i,k}(n)$ is always non-negative. Otherwise, we have
$$\Phi_{i,k}(n) \geq \alpha^{\Lambda^1_{i,n}(k) - 2\alpha} \geq \alpha^\alpha (\Lambda_{i,j}(k) - 3\alpha),$$ where the first inequlaity follows from Property 1. So in either case, 
(\ref{52}) holds. 

By combining (\ref{51-3}), (\ref{52}), Corollary~\ref{coro:phi-end},  and recalling $\alpha = \frac{10 \log d}{\log \log d}$, we have
 	$$V_2 \leq d^2 \sum_i \sum_{k} (\Lambda^1_i (k) - 3\alpha)_+ \leq \frac{d^2}{\alpha^\alpha}  \sum_i \sum_{k} \Phi_{i,k}(n) \leq   \frac{d^3}{\alpha^\alpha} m \leq \frac{m}{d}.\qedhere$$
\end{proof}

By Lemma~\ref{lem:vol-effect}, we have $\max_{i \in M_2, k \in [d]} \Lambda^2_i(k)  \leq \frac{1}{m} V_2 + 1 = O(1)$. Thus, we have shown that each of the two deterministic procedures yields a makespan of $O(\alpha) = O(\log d / \log \log d)$, thereby proving
the upper bound.

\medskip
\subsubsection{Algorithm for \vslpi}
\label{subsec:vslpi} 
We now give our $O( ( \log d /\log \log d)^{\frac{r-1}{r}})$-competitive algorithm for \vslpi.  Throughout the section, 
let $A$ denote the $O(\log d/ \log \log d)$-competitive algorithm for \vsmaxi defined in Section \ref{subsec:vsmaxi-derand}. 
Our \vslpi algorithm essentially works by using $A$ as a black box; however, we will perform a smoothing transformation on large loads 
before scheduling jobs with $A$. 

\smallskip
\noindent {\bf Algorithm Definition:}  We will apply the following transformation to all jobs $j$ that arrive online after Transformation 1 has been performed
(note that this is in replacement of Transformations 2 and 3 defined in Section \ref{subsec:vsmaxi-rand}).

\smallskip
{\bf  $\bullet$ Transformation 2:} If $p_j(k) > 1$, we reduce $p_j(k)$ to be 1. If this load reduction is applied 
in dimension $k$ for job $j$, we say $j$ is {\em large} in $k$; otherwise, $j$ is 
{\em small} in dimension $k$.
\smallskip

It is straightforward to see that Transformations 1 and 2 provide the following two properties: 

\begin{itemize} 
\item {\bf Property 1:}  $\sum_{j\in J} p_j(k) \leq m$ for all $k \in [d]$.
\item {\bf Property 2:} $0 \leq p_j(k) \leq 1$ for all $j \in J, k \in [d]$. 
\end{itemize} 

On this transformed instance, our algorithm simply schedules jobs using our \vsmaxi algorithm $A$.

\smallskip
\noindent {\bf Algorithm Analysis:} Let $\alpha = O(\log d / \log \log d$) be the competitive ratio of algorithm $A$. 
Clearly if we can establish $\alpha^{(r-1)/r}$-competitiveness for the scaled instance (i.e. just applying Transformation 1 to all jobs but not Transformation 2),
 then our algorithm is competitive on the original instance as well. Let {\sc OPT}$'(k,r)$ be the cost of the optimal solution on the scaled loads in 
dimension $k$. In Lemma \ref{lem:lpall-optlb}, we establish two lower bounds on {\sc OPT}$'(k,r)^r$.  

\begin{lemma}
	\label{lem:lpall-optlb}
$\displaystyle
\text{{\sc OPT}}'(k,r)^r \geq
		\max\left(\sum_{j\in J} p_j(k)^r, m\cdot\left(\sum_{j\in J} p_j(k) / m\right)^{r} \right)
		= \max\left(\sum_{j\in J} \left(p_j(k)^{r}\right), m \right).$
\end{lemma}
\begin{proof}
	Consider any fixed assignment of jobs, and let $J'(i) \subseteq J$ be the set of jobs assigned to machine $i$. Consider any fixed $k$. 
The first lower bound (within the max in the statement of the lemma) follows since 
\begin{equation*} 
\sum_{i \in M} \left(\sum_{j \in J'(i)} p_j(k)\right)^r \geq \sum_{i \in M} \sum_{j \in J'(i)} p_j(k)^r =  \sum_{j \in J} p_j(k)^r. 
\end{equation*} 
The second lower bound is due to the convexity of $x^r$ when $r \geq 1$. 
\end{proof}

Let $J(i) \subseteq J$ be the set of jobs assigned to machine $i$ by the online algorithm. 
Let $\ell(i,k)$ and $s(i,k)$ be the set of jobs assigned to machine $i$ 
that are large and small in dimension $k$, respectively. For brevity, let $\sigma_\ell(i,k) = \sum_{j \in \ell(i,k)} p_j(k)$ and $\sigma_s(i,k) = \sum_{j \in s(i,k)} p_j(k)$. 
Observe that since algorithm $A$ is $\alpha$-competitive on an instance with both Properties 1 and 2, we obtain the following additional two properties for the algorithm's 
schedule:  

\begin{itemize} 
\item {\bf Property 3:} $|\ell(i,k)| \leq \alpha$ for all $i \in M, k \in [d]$.
\item {\bf Property 4:} $\sigma_s(i,k) \leq \alpha$  for all $i \in M, k \in [d]$.
\end{itemize}  

Using these additional properties, the next two lemmas will bound the contribution of both large and small loads to the objective; namely, 
we need to bound both $\sigma_\ell(i,k)^r$ and $\sum_i \sigma_s(i,k)^r$ in terms of $\alpha$.   Lemma \ref{lem:lpalllarge} 
provides this bound for large loads, while Lemma \ref{lem:lpallsmall} will be used to bound small loads.

\begin{lemma} 
\label{lem:lpalllarge} 
$\sigma_\ell(i,k)^r = \left(\sum_{j \in \ell(i,k)}p_j(k)\right)^r \leq \alpha^{r-1} \sum_{j \in \ell(i,k)} p_j(k)^r. $ 
\end{lemma}
\begin{proof}  Let $h = |\ell(i,k)|$. 
Then, it follows that 
\begin{alignat*}{2} 
\left(\sum_{j \in \ell(i,k)}p_j(k)\right)^r & = \left(\frac{1}{h}\sum_{j \in \ell(i,k)} (p_j(k)\cdot h) \right)^r  \leq  \frac{1}{h}\sum_{j \in \ell(i,k)} (p_j(k)\cdot h)^r && \hspace{5mm} \text{(due to the convexity of $x^r$)} \notag\\ 
              & = h^{r-1} \sum_{j \in \ell(i,k)} p_j(k)^r  \leq \alpha^{r-1} \sum_{j \in \ell(i,k)} p_j(k)^r && \hspace{5mm}
                     \text{(by Property 3)}.\qedhere
\end{alignat*} 
\end{proof} 

Recall that by Property 1, we have that $\sigma_s(i,k) \leq m$.  Using this fact and along with Property 4, the general statement shown in Lemma
 \ref{lem:lpallsmall} will immediately provide us with the desired bound on $\sum_i \sigma_s(i,k)^r$ (stated formally in Corollary \ref{coro:small-identical}). 

\begin{lemma}
\label{lem:lpallsmall} 
	\label{lem:small-identical}
Let $f(x) =  x^r$ for some $r \geq 1$ whose domain is  defined over a set of variables $x_1, \ldots, x_n \in [0, \alpha]$ where $\alpha \geq 1$.
If $\sum_{i=1}^m x_i \leq m$, then
\begin{equation*}
\sum_{i = 1}^m f(x_i) \leq 2 m \ \alpha^{r -1}. 
\end{equation*} 
\end{lemma}
\begin{proof}
Let $\tilde{f} = \sum_{i = 1}^m f(x_i)$. We claim that $\tilde{f}$ is maximized when $0 < x_i < \alpha$ for at most one $i \in [m]$.
If there are two such variables $x_i$ and $x_j$ with $0 < x_i \leq x_j < \alpha$,
it is easy to see that we can further increase $\tilde{f}$ by decreasing $x_i$ and increasing $x_j$ by an infinitesimal equal amount
(i.e. $x_i \leftarrow x_i - \epsilon$ and $x_j \leftarrow x_j + \epsilon$) due to convexity of $f$. 

Hence, the $\tilde{f}$ is maximized when the multi-set $\{x_i : i \in [m]\}$ has $\lfloor m / \alpha  \rfloor$ copies of $\alpha$,
and one copy of $m - \alpha \lfloor m / \alpha  \rfloor$ (which is at most $\alpha$),
which gives, 
\begin{equation}
	\label{eqn:25}
	\tilde{f} \leq \lfloor m / \alpha  \rfloor f(\alpha) + f(m - \alpha \lfloor m / \alpha  \rfloor). 
\end{equation}

If $\lfloor m / \alpha  \rfloor \geq 1$, then it follows that 
\begin{alignat*}{2}  
\sum_{i = 1}^m f(x_i) = \tilde{f} &\leq (\lfloor m / \alpha  \rfloor + 1) f(\alpha) && \hspace{10mm}
                     \text{(by Eqn.\ \eqref{eqn:25} and since $m - \alpha \lfloor m / \alpha \rfloor \leq \alpha)$} \\ 
                  &\leq  2 (m/ \alpha) \alpha^r  =  2 m \ \alpha^{r -1} && \hspace{10mm}
                      \text{(since $\lfloor m / \alpha  \rfloor \geq 1$).} 
\end{alignat*} 

In the case where $m < \alpha$, $\tilde{f}$ is maximized by making single $x_i = m$. Therefore
$\tilde{f} \leq f(m)  = m^r \leq  m  \alpha^{r -1}$. 
\end{proof}

\begin{corollary}
	\label{coro:small-identical}
	For all dimensions $k$, $\sum_{i \in M} \sigma_s(i,k)^r \leq 2 m \ \alpha^{r -1}$. 
\end{corollary}

%
%
%

We are now ready to bound $\|\Lambda(k)\|_r$ against $\text{{\sc OPT}}'(k, r)$.
\begin{lemma} 
\label{lem:lpall-algobound}
For all dimensions $k$, $\|\Lambda(k)\|_r = O(\alpha^{(r-1)/r}) \text{{\sc OPT}}'(k, r)$, i.e., the $L_r$ norm of the vector load is at most $O(\alpha^{(r-1)/r})$ 
times the $L_r$ norm of the vector load of the optimal solution. 
\end{lemma} 
\begin{proof}
Using Lemmas \ref{lem:lpall-optlb}, \ref{lem:lpalllarge}, and Corollary~\ref{coro:small-identical}, we have the following bound for $|\Lambda(k)\|_r^r = \sum_{i \in M} \left(\sum_{j \in J(i)} p_j(k)\right)^r$: 

\begin{alignat*}{2} 
	 \sum_{i \in M} \left(\sum_{j \in J(i)} p_j(k)\right)^r \notag 
             &= \sum_{i \in M} \left(\sum_{j \in \ell(i,k)} p_j(k) + \sum_{j \in s(i,k)} p_j(k) \right)^r \notag \\
             & \leq \sum_{i \in M}\left(2\max\left(\sum_{j \in \ell(i,k)} p_j(k), \sum_{j \in s(i,k)} p_j(k)\right)\right)^r \notag\\
             & \leq 2^r \sum_{i \in M} \left(\left(\sum_{j \in \ell(i,k)}p_j(k)\right)^r + \left(\sum_{j \in s(i,k)} p_j(k)\right)^r \right) \notag \\
             & \leq 2^r \left(\alpha^{r-1} \sum_{j \in \ell(i,k)} p_j(k)^r + 2m\cdot \alpha^{r-1} \right) 
                 && \hspace{5mm}  \text{(by Lemma \ref{lem:lpalllarge} and Corollary~\ref{coro:small-identical})} \\
             & \leq (2^{r+1} \alpha^{r-1})\text{{\sc OPT}}'(k, r)^r
                 && \hspace{5mm} \text{(by Lemma~\ref{lem:lpall-optlb})} 
\end{alignat*}  
which, raising both the LHS and RHS to $1/r$, gives us 
\begin{equation} 
\|\Lambda(k)\|_r \leq \left(2^{1 + 1/r} \alpha^{(r-1)/r}\right)\text{{\sc OPT}}'(k, r) = O(\alpha^{(r-1)/r}) \text{{\sc OPT}}'(k, r) .  \notag
\end{equation} 

\end{proof} 

The upper bound in Theorem \ref{thm:vsall} now follows immediately from Lemma \ref{lem:lpall-algobound}. 

}

\section{Unrelated Machines}
\label{sec:unrelated} 

Now, we consider the online vector scheduling problem for unrelated machines. 
\full{
In this section, we obtain tight upper and lower bounds for this problem, 
both for the makespan norm (Theorem~\ref{thm:vsmaxu}) and for arbitrary $L_r$ norms
(Theorem~\ref{thm:vsany}).
}
\submit{
\hspace{-3mm}
We give tight upper and lower bounds for this problem
both for the makespan norm (Theorem~\ref{thm:vsmaxu}) and for arbitrary $L_r$ norms
(Theorem~\ref{thm:vsany}). In the extended abstract, we only present the upper 
bounds (for general $L_r$ norms; the makespan result was known~\cite{MeyersonRT13});
the lower bounds appear in the full paper.
}

\full{
\subsection{Lower Bound for \vslpu} 
\label{app:unlb}

In this section we prove the lower bound in Theorem~\ref{thm:vsany}, 
i.e., we show that we can force any algorithm to make an assignment 
where there exists a dimension $k$ that has cost at least $\Omega(\log d + r_k)$ where $1 \leq r_k  \leq \log m$.

Our construction is an adaptation of the lower bounds in \cite{Caragiannis08} and \cite{AwerbuchAGKKV95} but
for a multidimensional setting.   Informally, the instance is defined as follows. We set $m = d$ and then associate $i$th machine with the $i$th dimension, 
i.e., machine $i$ only receives load in the $i$th dimension.
We then issue jobs in a series of $\log d +1$ phases.  In a given phase,
there will be a current set of {\em active machines}, which are the only machines that  can be loaded in the current phase and for 
the rest of the instance (so once a machine is inactivated it stays inactive). More specifically, in a given phase we 
arbitrarily pair off the active machines and then issue one job for each pair, 
where each job has unit load but is defined such that it must be assigned to a unique machine in its pair. When a phase completes, we inactivate all the machines that did not receive load (so
we cut the number of active machines in half). This process eventually produces a load of $\log d +1 $ on some machine, whereas reversing the decisions of the 
algorithm gives an optimal schedule where $L_k  = 1$ for all $k \in [d]$.


More formally let $d = 2^h$. The adversary sets the instance target parameters to be $T_k = 1$ for all $k \in [d]$ (it will be clear from our construction that these targets are feasible). 
For each job $j$,  let $m_{1}(j), m_{2}(j) \in [m]$ denote the machine pair the adversary associates with job $j$.  We define $j$ to have unit load on machines $m_{1}(j)$, $m_{2}(j)$ in their respective dimensions and 
arbitrarily large load on all other machines. Formally, $p_{i,j}(k)$ is defined to be  

\begin{equation*} 
p_{i,j}(k) = \left\{ 
\begin{array}{ll}
0 & \text{if $i \neq k$ and $i \in \{m_{1}(j), m_{2}(j)\}$}\\
1 & \text{if $i = k$ and  $i \in \{m_{1}(j), m_{2}(j)\}$}\\ 
\infty & \text{otherwise}. 
\end{array}\right.
\end{equation*}


As discussed above, the adversary issues jobs in $h+1$ phases. Phases $1$ through $h$ will work as previously specified (we describe how the final $(h+1)$th phase works shortly).  Let $S_\ell$ denote the active machines in phase $\ell$. In the $\ell$th phase, we issue a set of jobs $J_\ell$ where $|J_\ell| = 2^{h-\ell}$. We then pair off the machines in $S_\ell$ and use each machine pair as $m_{1}(j)$ and $m_{2}(j)$ for a unique job $j \in J_\ell$.  Clearly the algorithm must schedule $j$ on  $m_{1}(j)$ or $m_{2}(j)$, and thus  $2^{h-\ell}$ machines accumulate an additional load of 1 in phase $\ell$.   Machines that receive jobs in phase $\ell$ remain active in phase $\ell +1$; all other machines are set to be inactive. In the final phase $h+1$, there will be a single remaining active machine $i'$; thus, we issue a single job $j'$ with unit load that must be scheduled on $i'$ (note that this final phase is added to the instance only to make our target vector feasible). 
 
Based on this construction, there will exist a dimension $k'$ at the end of the instance that has load $h+1$ on machine $k'$ and 0 on all other machines. Observe that  the optimal schedule is obtained by reversing the decisions of the algorithm, which places a unit load on one machine in each dimension. Namely, if $j$ was assigned to $m_{1}(j)$, then the optimal schedule assigns $j$ to $m_{2}(j)$ (and vice versa), with the exception that $j'$ is assigned to its only feasible machine. 

In the case that $\log d \geq r_{k'}$, the adversary stops.  Since $T_{k'} = 1$ and $L_{k'} = h +1 = \log d + 1$, we have that $L_{k'}/T_{k'}= \Omega(\log d + r_k)$. 
If $\log d < r_{k'}$, then the adversary stops the current instance and begins a new instance.   In the new instance, we simply simulate the lower bound from \cite{AwerbuchAGKKV95} in dimension $k'$ (i.e., the only dimension that receives load is dimension $k'$; the adversary also resets the target vectors accordingly).  Here, the adversary forces the algorithm to be $\Omega(r_{k'})$-competitive, which, since $\log d < r_{k'}$, gives us the desired bound of $\Omega(\log d + r_{k'})$. 

}
\full{
\subsection{Upper Bound} 
\label{subsec:lpnorm} 
}
Our goal is to prove the upper bound in Theorem~\ref{thm:vsany}. 
Recall that we are given targets $T_k$, and we have to show that
$\|\Lambda(k)\|_{r_k} = O(\log d + r_k) \cdot T_k$ 
for all $k \in [d]$. ($\Lambda(k)$ is the load vector in
dimension $k$ and $r_k$ is the norm that we are optimizing.) 
First, we normalize $p_{i,j}(k)$ to $p_{i,j}(k)/T_k$ for all dimensions $k$; 
to keep the notation simple, we will also denote this normalized load $p_{i,j}(k)$.  
This ensures that the target objective is 1 in every dimension. (We assume wlog 
that $T_k > 0$. If $T_k=0$, the algorithm discards all assignments that put 
non-zero load on dimension $k$).  



\full{
\subsubsection{Description of the Algorithm}
}
\submit{
\subsection{Description of the Algorithm}
}
 
As described in the introduction, our algorithm
is greedy with respect to a potential function defined on
modified $L_{r_k}$ norms. 
Let $L_{k} = \|\Lambda(k)\|_{r_k}$ denote
the $L_{r_k}$-norm of the machine loads 
in the $k$th dimension,
and $q_k = r_k + \log d$ denote the
desired competitive ratio; all logs are base 2.   
We define the potential for dimension $k$ as
$\Phi_k = L_k^{q_k}$. The potentials for the $d$
different dimensions are combined using a weighted linear 
combination, where the weight of dimension $k$ is
$\alpha_k  = (3 q_k)^{-q_k}$. 
Note that dimensions that 
allow a smaller slack in the competitive ratio 
are given a larger weight in the potential. 
We denote the combined potential by $\Phi = \sum_{k = 1}^d \alpha_k \cdot \Phi_k$. 
The algorithm
assigns job $j$ to the machine that minimizes the increase
in potential $\Phi$. 

\full{
\subsubsection{Competitive analysis} 
}
\submit{
\subsection{Competitive analysis of the Algorithm} 
}
Let us fix a solution satisfying the target objectives, and call  
it the optimal solution. 
Let $ \Lambda_i(k)$ and $\Lambda_i^*(k)$ be the 
load on the $i$th machine in the $k$th dimension for the 
algorithmic solution and the optimal solution respectively.
We also use $L^*_k$ to denote the $L_{r_k}$ norm in the 
$k$th dimension for the optimal solution; we have already
asserted that by scaling, $L^*_k = 1$.

Similar to \cite{AspnesAFPW97,Caragiannis08}, we compare the 
actual assignment made by the algorithm (starting with zero load on every 
machine in every dimension) to a hypothetical assignment made by the optimal solution 
starting with the final algorithmic load on every machine 
(i.e., load of $\Lambda_i(k)$ on machine $i$ in dimension $k$). 

We will need the following fact for our analysis, which follows by observing that all parameters are positive, the function is continuous in the domain, and its derivative is non-negative. 

\begin{fact}  
\label{fact:inc-fact} 
The function $f(x_1, x_2,$ $\ldots, x_m) = \left(\sum_i (x_i+ a_i)^w\right)^z-  \left(\sum_i x_i^w\right)^z$
is non-decreasing if for all $i \in [m]$ we restrict the domain of $x_i$ to be $[0, \infty)$, $w \geq 1, z \geq 1$, and $a_i \geq 0$.
\end{fact} 

Using greediness
of the algorithm and convexity of the potential function, we 
argue in Lemma \ref{lem:potbound}
that the change in potential in the former process is upper
bounded by that in the latter process.
\begin{lemma} 
\label{lem:potbound} 
The total change in potential in the online algorithm satisfies:\\
\begin{equation*} 
\sum_{k=1}^d \alpha_k L_k^{q_k}   = \Phi(n) - \Phi(0)
\leq \sum_{k=1}^d  \alpha_k \Big(\sum_{i=1}^m \Big(\Lambda_{i}(k) + \Lambda_{i}^*(k) \Big)^{r_k}\Big)^{q_k/r_k}  - \sum_{k=1}^d \alpha_k L_k^{q_k}
\end{equation*}  
\end{lemma} 

\begin{proof}
Let $y_{i,j} = 1$ if the algorithm assigns job $j$ to machine $i$; otherwise, $y_{i,j} =0$.  Define $y_{i,j}^*$ similarly but for the optimal solution's assignments. 
We can express the resulting change in potential from scheduling job $j$ as follows. 

\begin{eqnarray}
\Phi(j) - \Phi(j-1) &= \sum_{k=1}^d \alpha_k\left(L_k^{q_k}(j) - L_{k}^{q_k}(j-1) \right) = \sum_{k=1}^d \alpha_k \Big( \Big(\sum_{i=1}^m \Lambda_{i,j}^{r_k}(k) \Big)^{q_k/r_k} - \Big(\sum_{i=1}^m\Lambda_{i,j-1}^{r_k}(k)\Big)^{q_k/r_k}\Big) \nonumber\\ 
& = \sum_{k=1}^d \alpha_k \Big( \Big(\sum_{i=1}^m \Big(\Lambda_{i,j-1}(k) + p_{i,j}(k)\cdot y_{i,j} \Big)^{r_k}\Big)^{q_k/r_k}  - \Big(\sum_{i=1}^m\Lambda_{i,j-1}^{r_k}(k)\Big)^{q_k/r_k} \Big). \label{eq:pot1} \vspace{-2mm}
\end{eqnarray}

%
Since the online algorithm schedules greedily based on $\Phi(j)$, using optimal schedule's assignment for job $j$
must result in a potential increase that is at least as large. Therefore by \eqref{eq:pot1} we have 

\begin{equation} 
\label{eq:pot2} 
\Phi(j) - \Phi(j-1) \leq  \sum_{k=1}^d \alpha_k\Big( \Big(\sum_{i=1}^m \Big(\Lambda_{i,j-1}(k) + p_{i,j}(k)\cdot y^*_{i,j} \Big)^{r_k}\Big)^{q_k/r_k}  
                    - \Big(\sum_{i=1}^m\Lambda_{i,j-1}^{r_k}(k)\Big)^{q_k/r_k} \Big).\\ 
\end{equation} 

\noindent
 As loads are non-decreasing, $\Lambda_{i}(k) \geq \Lambda_{i,j-1}(k)$. Also note that $r_k \geq 1$ and $$q_k/r_k  = (r_k + \log d)/r_k > 1.$$
Thus, we can apply Fact \ref{fact:inc-fact} to \eqref{eq:pot2} (setting $w = r_k$, $z = q_k/r_k$, and $a_i =  p_{i,j}(k)\cdot y^*_{i,j}$) to obtain
\begin{equation} 
\Phi(j) - \Phi(j-1)  \leq \sum_{k=1}^d \alpha_k \Big( \Big(\sum_{i=1}^m \Big(\Lambda_{i}(k) + p_{i,j}(k)\cdot y^*_{i,j}  \Big)^{r_k} \Big)^{q_k/r_k} 
                    - \Big(\sum_{i=1}^m\Lambda_{i}^{r_k}(k)\Big)^{q_k/r_k}  \Big). \\  
\end{equation} 

\noindent
We can again use Fact \ref{fact:inc-fact} to further bound the potential increase (using the same values of $a_i$, $w$, and $z$, but now $ x_i = \Lambda_{i,j-1}^*(k)$): 
\begin{alignat}{2}
\Phi(j) - \Phi(j-1) 
             & \leq \sum_{k=1}^d \alpha_k \Big( \Big(\sum_{i=1}^m \Big(\Lambda_{i}(k) + \Lambda^*_{i,j-1}(k)
                    + p_{i,j}(k)\cdot y^*_{ij} \Big)^{r_k}  \Big)^{q_k/r_k} 
                   - \Big(\sum_{i=1}^m \Big(\Lambda_{i}(k) + \Lambda_{i,j-1}^*(k) \Big)^{r_k} \Big)^{q_k/r_k}  \Big) \notag \\ 
             & = \sum_{k=1}^d \alpha_k \Big( \Big(\sum_{i=1}^m \Big(\Lambda_{i}(k)+ \Lambda_{i,j}^*(k)  \Big)^{r_k}\Big)^{q_k/r_k} 
                    - \Big(\sum_{i=1}^m \Big( \Lambda_{i}(k) + \Lambda_{i,j-1}^*(k) \Big)^{r_k} \Big)^{q_k/r_k}  \Big).   \label{eq:pot3}
\end{alignat}

\noindent Observe that for a fixed $k$, the RHS of \eqref{eq:pot3} is a telescoping series if we sum 
over all jobs $j$: 
{\small
\begin{eqnarray*}
& & \sum_{j=1}^n \alpha_k \Big( \Big(\sum_{i=1}^m \Big(\Lambda_{i}(k)+ \Lambda_{i,j}^*(k)  \Big)^{r_k}\Big)^{q_k/r_k} 
                    - \Big(\sum_{i=1}^m \Big( \Lambda_{i}(k) + \Lambda_{i,j-1}^*(k) \Big)^{r_k} \Big)^{q_k/r_k}  \Big)  \\
                    & = &    \alpha_k \Big(\sum_{i=1}^m \Big(\Lambda_{i}(k) + \Lambda_{i}^*(k) \Big)^{r_k}\Big)^{q_k/r_k}  
                       - \Big(\sum_{i=1}^m \Big(\Lambda_{i}(k) \Big)^{r_k}\Big)^{q_k/r_k}. \label{eq:pot4}
\end{eqnarray*}
}
We have $$\sum_{j=1}^n(\Phi(j) - \Phi(j-1)) = \Phi(n) - \Phi(0),$$ since this is also a telescoping series. 
By definition, $\Phi(0) = 0$ and $\Phi(n) = \sum_{k=1}^d \alpha_k L_{k}^{q_k}$.  Using these facts along with \eqref{eq:pot3} and \eqref{eq:pot4}, we establish the lemma: 

\begin{eqnarray*} 
\sum_{k=1}^d \alpha_k L_k^{q_k} 
& = & \sum_{j=1}^n(\Phi(j) - \Phi(j-1)) \hspace{5mm} \text{(since $\Phi$ telescopes, $\Phi(0) = 0$, and $\Phi(n) = \sum_{k=1}^d \alpha_k L_{k}^{q_k}$)} \\    
& \leq & \sum_{k=1}^d  \alpha_k \Big(\sum_{i=1}^m \Big(\Lambda_{i}(k) + \Lambda_{i}^*(k) \Big)^{r_k}\Big)^{q_k/r_k} 
                      - \sum_{k=1}^d \alpha_k L_k^{q_k} \Big)   \hspace{5mm} \text{(by \eqref{eq:pot3} and  \eqref{eq:pot4})}.\qedhere 
\end{eqnarray*} 

\end{proof}

We proceed by applying Minkowski inequality (e.g.,~\cite{wiki:MinkowskiInequality}), 
which states that for any two vectors {\bf $v_1$} and {\bf $v_2$}, we have 
$\|v_1 +  v_2\|_r \leq \|v_1\|_r+ \|v_2\|_r$. Applying this inequality 
to the RHS in Lemma~\ref{lem:potbound}, we obtain
\begin{alignat}{2} 
\sum_{k=1}^d \alpha_k L_k^{q_k}   & \leq \sum_{k=1}^d  \alpha_k\Big(\Big(\sum_{i=1}^m \Lambda_{i}^{r_k}(k)\Big)^{1/r_k}
                        + \Big(\sum_{i=1}^m( \Lambda_{i}^*(k))^{r_k} \Big)^{1/r_k}\Big)^{q_k} 
                       - \sum_{k=1}^d \alpha_k L_k^{q_k}   \notag \\ 
                 &= \sum_{k=1}^d \alpha_k \Big( L_k + L_k^*\Big)^{q_k}  - \sum_{k=1}^d \alpha_k L_k^{q_k} .  \label{eq:lp5}                            
\end{alignat} 

Next, we prove a simple lemma that we will apply to inequality  \eqref{eq:lp5}.
\begin{lemma}
	\label{lem:15}
	$(L_k + L_k^*)^{q_k} \leq e^{1/2} L_k^{q_k} + (3q_k\cdot L_k^*)^{q_k}$ for all $k \in [d]$. 
\end{lemma}

\begin{proof} 
First consider the case  $L_k \geq 2q_k \cdot L_k^*$.  Then it follows,
\begin{alignat}{2} 
(L_k + L_k^*)^{q_k} & \leq \left(1 + 1/(2q_k)\right)^{q_k} \cdot L_k^{q_k} \notag \\ 
                                & \leq \left(e^{1/(2q_k)}\right)^{q_k} \cdot L_k^{q_k} = e^{1/2} L_k^{q_k}.
\end{alignat} 
Otherwise $L_k < 2q_k \cdot L_k^*$, and then we have $$(L_k + L_k^*)^{q_k} \leq (3q_k\cdot L_k^*)^{q_k}.$$ 
Combining these two upper bounds completes the proof. 
\end{proof} 
%
%

Thus, we can rearrange \eqref{eq:lp5} and bound $2\sum_{k=1}^d \alpha_k L_{k}^{q_k}$ as follows:
%

\begin{alignat}{2} 
2\sum_{k=1}^d \alpha_k L_{k}^{q_k}   & \leq  \sum_{k=1}^d \alpha_k \left( L_k + L_k^*\right)^{q_k}   \leq e^{1/2} \sum_{k=1}^d \alpha_k L_k^{q_k} +  \sum_{k=1}^d \alpha_k  (3q_k \cdot L_k^*)^{q_k} && \hspace{5mm} \text{(by Lemma \ref{lem:15})} \notag \\
         &= e^{1/2} \sum_{k=1}^d \alpha_k L_k^{q_k} +  \sum_{k=1}^d  (L_k^*)^{q_k} \label{eq:lp8} .
\end{alignat}
Note that the last equality is due to the fact that $\alpha_k^{-1}  = (3q_k)^{q_k}$. 
By our initial scaling, $L_k^* = 1$ for all $k$. Therefore, after rearranging \eqref{eq:lp8}, we obtain 
%
\begin{equation*}
\left(2-e^{1/2}\right)\sum_{k=1}^d \alpha_k L_{k}^{q_k} \leq  \sum_{k=1}^d  (L_k^*)^{q_k} \leq d, 
\end{equation*} 
which for any fixed $k$ implies 
%

\begin{alignat*}{2} 
L_{k} & 
\leq \frac{1}{\left(2-e^{1/2}\right)^{1/{q_k}}} \cdot  \left(\frac{d}{\alpha_k} \right)^{1/{q_k}}
\leq \frac{1}{2-e^{1/2} }\cdot  \left(\frac{d}{\alpha_k} \right)^{1/{q_k}} \\
& = \frac{3}{2-e^{1/2}} \cdot  \left(d^{\frac{1}{r_k + \log d}}\right)q_k   
 <  10 \cdot  d^{\frac{1}{\log d}} \cdot q_k = 20 q_k = O(r_k + d),
\end{alignat*}
where the first inequality uses $q_k \geq 1$ and $2 - e^{1/2} < 1$. This completes the proof of the upper bound claimed in Theorem~\ref{thm:vsany}.

\section*{Acknowledgements} 
S. Im is supported in part by NSF Award CCF-1409130. A part of this work was done by J. Kulkarni
at Duke University, supported in part by NSF Awards CCF-0745761,
CCF-1008065, and CCF-1348696. N. Kell and D. Panigrahi are supported in part by NSF Award CCF-1527084, 
a Google Faculty Research Award, and a Yahoo FREP Award. 

\bibliographystyle{plain}
\bibliography{ref}





\end{document}